# Strategic play in stable marriage problem

(Mircea Digulescu)
14.05.2016 – 24.08.2016

## ABSTRACT

The stable marriage problem, as addressed by Gale and Shapely [1] consists of providing a bipartite matching between n "boys" and n "girls" - each of whom have a totally ordered preference list over the other set - such that there exists no "boy" and no "girl" that would prefer each other over their partner in the matching. In this paper, we analyze the cases of strategic play by the "boys" in the game directly inspired by this problem. We provide an $O(n^3)$ algorithm for determining a matching which is not necessarily stable in the Gale-Shapely sense, but it is coalition-stable, in that no player has a selfish interest to leave the resulting grand coalition to join any potential alternative one which might feasibly form, and is also man-optimal. Thus, under a realistic assumption set, no player has an interest to "destabilize" the matching, even though he theoretically could. The resulting matching is often better than the naïve Gale-Shapely one for *some* (not all) of the "boys", being no worse for the rest. This matching is more realistic (stable) than the one produced by top-trading-cycles method, thus offering a qualitative improvement over the latter. Furthermore, we analyze the situation when players are allowed to make strategic threats (i.e. be willing to sacrifice their own outcome to hurt others), offer a relevant example to illustrate the benefits of this form of play, and ultimately provide an exponential time algorithm which tries to determine a good threat-making strategy. We then briefly examine a few other non-conventional possibilities a player has to affect his outcome. Most common variations to the game model are also described and analyzed with regard to applicability of the methods in this paper. Finally, a few examples of real-life problems which can be modeled and solved with the methods in this paper are presented.

## 1. OVERVIEW AND MOTIVATION

*Introduction*

The statement of the marriage problem consists of the following:
- A set of **n** elements **B**, of "boys" (suitors, proposers, "buyers", etc.) who are interesting in obtaining a match with exactly one of the girls in set **G**.
- A set of **n** elements **G**, of "girls" (acceptors, "sellers", etc.) who are interesting in obtaining a match with exactly one of the boys in set **B**.
- For each boy $b_i$ in **B**, a permutation $Pb_i$ of the girls in **G**, representing his preference

list over the girls. A girl which appears earlier in **Pb$_i$** is preferred by boy **b$_i$** over one that appears later.
- For each girl **g$_j$** in **G**, a permutation **Pg$_j$** of the boys in **B**, representing her preference list over the boys. A boy which appears earlier in **Pg$_j$** is preferred by girl **g$_j$** over one that appears later.

By convention, for brevity, I will refer to an element of **B** as boy **b$_i$** and to an element of **G** as girl **g$_j$**, in general.

An interesting case (which is the only one analyzed in this paper) is that when the matching is the result of a game played as follows:
- Every unmatched boy **b$_i$** proposes to some girl **g$_i$** in his preference list.
- Every girl **g** in the game always keeps her best match (the highest on her preference list) among all proposals she received and dumps all the rest (including in some cases her former partner).
- When there are no more unmatched boys, the game ends and the resulting matching is the final matching.

**Example 1.1:**
B = {b1, b2, b3, b4, b5, b6, b7}
G = {g1, g2, g3, g4, g5, g6, g7}

Preference of boys, in descending order:

| Pb1 | **g1**, g2, g3, g4, g5, g6, g7 |
|---|---|
| Pb2 | **g2, g4**, g1, g3, g5, g6, g7 |
| Pb3 | **g2, g3**, g1, g4, g5, g6, g7 |
| Pb4 | **g1, g2, g7, g6, g5**, g3, g4 |
| Pb5 | **g3, g2**, g1, g4, g5, g6, g7 |
| Pb6 | **g6, g5**, g1, g2, g3, g4, g7 |
| Pb7 | **g4, g7**, g1, g2, g3, g5, g6 |

Preference of girls, in descending order:

| Pg1 | b1, b2, b3, b4, b5, b6, b7 |
|---|---|
| Pg2 | b5, b2, b3, b4, b1, b6, b7 |
| Pg3 | b3, b5, b1, b2, b4, b6, b7 |
| Pg4 | b1, b2, b7, b3, b4, b5, b6 |
| Pg5 | b1, b2, b3, b4, b5, b6, b7 |
| Pg6 | b6, b4, b1, b2, b3, b4, b5 |
| Pg7 | b1, b2, b3, b5, b6, b7, b4 |

|
A solving (or scenario) of the stable marriage problem consists of a bipartite matching between the elements of **B** and the elements of **G**. A partial solving (partial scenario) consists of a partial matching, with some elements remaining unmatched.

It is interesting to analyze matchings which have some sort of stability property. That is, boys will not dump their current partners (in the solving) – and thus become unmatched – so that they can then propose to another girl (or girls) in the hope they eventually end up better off. Furthermore, the girls must not receive better offers so as to jeopardize the matching of their partner boy, leaving him unmatched again. It is reasonably conceived that a matching without these kind of stability properties is not a realistic solution, since actors within it will have no rational interest to "stick to it" (in the problem's current formulation) and thus will continue the game resulting in a new matching. Cases of incomplete information need to be analyzed separately.

**Definition 1.2 (strict/Gale-Shapely stability)**
A (complete) solving of a marriage problem instance such that there exists no tuple ($b_i$, $g_j$) such that $b_i$ ranks higher in $g_j$'s preference lists then her partner in the solving and $g_j$ ranks higher in $b_i$'s preference lists then his partner in the solving, is called ***a strictly stable solution***, or Gale-Shapley-stable. |

**Example 1.3:**
For the statement in Example 1.1, the following would be a stable solution.
S = b1-g1, b2-g4, b3-g3, b4-g5, b5-g2, b6-g6, b7-g7.|

**Definition 1.4 (coalition stability)**
A (complete) solving of a marriage problem instance is such that there exists no coalition **C** = {$b_{i1}$, $b_{i2}$, … , $b_{ik}$} of boys such that should all of them decide to dump their current partners (and thus become unmatched) and then collude in executing a strategy of proposing to girls, at least one of them (a boy in C) will be better of with the resulting partner <u>and the rest **of those in C** will be no-worse off</u>, is called a ***coalition-stable solution***. |
*Discussion*: Note that players outside C might also react, so coalition stability refers to the entire set B of boys.

The condition that all members of the coalition C are no-worse off is required for the common sense intuition that a boy cannot be convinced to become part of a coalition that generates a worse of result for him than if he did not join. However, this special case is possible for example in cases where preferences can be tied.

**Example 1.5:**
For the statement in Example 1.1, the following would be a coalition-stable solution.
S = b1-g1, b2-g4, b3-g2, b4-g5, b5-g3, b6-g6, b7-g7.|

The number of boys and the number of girls does not have to be exactly the same. In case there are more girls than boys, some girls will become matched with some "fictitious" boy index above the number of actual boys, signifying in effect that she remained unmatched. The same applies conversely if there are more boys than girls.

*Prior works*

The stable marriage problem was introduced to literature largely by Gale and Shapely in [1 - College Admissions and the Stability of Marriage, D. Gale and L. S. Shapley, 1962] where they provided an O($n^2$) algorithm which produces a stable matching between the two groups. A stable matching is one such that there exists no boy and no girl such that they both prefer each other over their respective partners in the matching. Robert W. Irving studied in [2 – Stable Marriage and indifference, Robert W. Irving, 1994] the case where the preferences lists allow indifference between different partners and provided several algorithms for determining stable matchings (if they exist) in such cases. A few fundamental issues concerning strategic play were studied in [3 - Machiavelli and the Gale-Shapley Algorithm, L. E. Dubins and D. A. Freedman, 1981] by Dubins and Freedman where they showed that no coalition of boys can improve the outcome *for all of them*, by lying about their preferences. In a sequel paper [4 - Ms. Machiavelli and the Stable Matching Problem, David Gale and Marilda Sotomayor, 1985], Gale and Sotomayor analyzed the case with lying by the girls. In [5 - Marriage, honesty, and stability, Nicole Immorlica, Mohammad Mahdian, 2005] the authors concerned themselves with lying by the boys in matching markets where one side only has a constant number of preferences and also acknowledged that "no matching mechanism based on a stable marriage algorithm can guarantee truthfulness as a dominant strategy for participants". Another prior paper concerning itself with misrepresentation is [6 - Misrepresentation and stability in the marriage problem, Alvin E Roth, 1984]. Some variations to the stable marriage problem concerning the simultaneous introduction of ties and seeking a more balanced matching (favoring the girls more than in Gale-Shapely) have been shown in [7 - Hard variants of stable marriage, David F Manlove, Kazuo Iwama, Shuichi Miyazaki, Yasufumi Morita, 2002] to be hard (with regard to NP completeness). A series of existential results concerning strategic play in stable marriage problem and some of its variations have been presented by Roth in [8 - The college admissions problem is not equivalent to the marriage problem, Alvin E Roth].

In [9 - Gale-Shapley Stable Marriage Problem Revisited: Strategic Issues and Applications, Chung-Piaw Teo, Jay Sethuraman,Wee-Peng Tan, 1999] the authors analyzed strategic play by girls and offered several results. Finally, in [10 - Cheating by Men in the Gale-Shapley Stable Matching Algorithm, Chien-Chung Huang, 2006], Huang analyzed cases of strategic plays by boys consisting of collusion in order to achieve a better outcome for *some* (not all) of them. The novelty was that such outcomes are not necessarily stable. They are however no worse than Gale-Shapely and are also on the Pareto frontier of outcomes no worse than Gale-Shapely. This prompted Huang to imply that such outcome is the best possible for boys. His approach consisted of improving a Gale-Shapely matching by discovering and materializing "trading cycles", using the top-trading-method described in [11 - Pareto Optimality in House Allocation Problems, David J. Abraham, Katarína Cechlárová, David F. Manlove, Kurt Mehlhorn, 2005].

A brief survey of results concerning stable marriage problem was published in 2008, in [12 - A Survey of the Stable Marriage Problem and Its Variants, Kazuo Iwama, Shuichi Miyazaki, 2008] by Iwama and Miyazaki.

A general and very relevant situation where the actual utility gained by men in a matching can be modeled in terms of transferable utility (e.g. monetary value) with side-payments allowed has been studied by Rahul Jain in [13 - Designing a strategic bipartite matching market, Rahul Jain, 2007].

*Overview of this paper*

In this paper we concern ourselves primarily with generating as good an outcome as realistically possible for some boy.

The main result of this paper is to provide a more realistic matching than the one by [10] for the case where boys are allowed to collude. Our matching more adequately accounts for the power and interest boys have to propose successfully to certain girls. In the terminology of [10], we claim that Huang's result is sometimes unrealistic because it entails cooperation of more accomplices than needed, with some accomplices actually having the incentive to refuse cooperation since they can obtain a better outcome (become cabalists in terminology of [10]) if other accomplices (which are same off in both cases) support them as well. We provide an illustrating example of such a situation in the beginning of Section 3. In that section we also provide an $O(n^3)$ algorithm to determine such a matching, which we call coalition stable, under the explicit assumption set of that section. We also prove that such a coalition stable matching is unique and optimal for all men under the Section's assumptions.

Another important contribution in this paper consists of an analysis in Section 4 of cases where boys are allowed to try to persuade others into a coalition with an outcome more favorable to them by threatening that in case of non-cooperation the latter's outcome will be degraded much more severely. For such threats to be both useful and credible, the player making them must be willing to risk getting a worse outcome than Gale-Shapely in case his ultimatums are rejected. We give examples showing a number of positive existential results, discuss complexities and traits of the ensuing model and progress towards solving the game by offering a super-exponential time algorithm to find a strategy which guarantees that some particular boy will be matched to a preference at least as good as a fixed one. The algorithm has the drawback of producing false negatives occasionally.

In Section 5, we discuss six further possibilities of non-conventional play which might affect the outcomes for some boy, namely adding/removing boys/girls, coopting players in coalitions without incentivizing them within the game and making use of drone players. We also show how such possibilities might become available to some player from a motivational perspective of the involved.

In Section 6, we discuss in reasonable detail six major variations to the game model, namely girls with more slots, indifference in the preference lists of boys, boys not knowing about all the girls in play, boys not knowing about all other boys, girls as players and imperfect information. We provide examples to illustrate the impact such variations have on prior results.

Finally, in Section 7 we show how four very relevant real-life games can be modeled as stable marriage problem instances and also present an adaptation to the stable marriage problem to allow for bids to be submitted by boys alongside proposals which can affect how preferred they are in the eyes of the girls.

*Motivation*

This paper is motivated primarily by the numerous economic situations which can be formulated as instances of the stable marriage problem. There are many examples in literature.

In Section 7, we discuss four real life games namely Renter-Landlord, Contractor – Project, Student – Program placement and Ads placement and analyze how they can be modeled as stable marriage problem instances. In the course of doing this we also provide an adaptation to stable marriage algorithms allowing a major variation pertaining to the contractor – project problem, under which boys are allowed to submit bids alongside the proposals which affect (but not solely determine) their rankings in the eyes of the girls. We also referenced a fifth game, namely Wireless Communications where our results also apply.

For almost all such situation strategic play can play an important role. Because of this, we consider important that the limitations of [10] (which is currently to the best of our knowledge the state of the art in this regard) are noted the qualitative improvements we make and the issues we discuss in this paper are taken into account.

## 2. PRELIMINARIES

In this section we proceed by introducing some preliminary notions, assumptions and basic results.

**Assumption 2.1 (girls are robotic players)**
In the framework of the first few sections, girls are considered passive (robotic-players)

who always accept the better choice between their then-current partner and the proposer. We will therefore use terminology player to refer strictly to some boy.|

For analysis purposes, we need to formalize how this "marriage game" is to be played out in real life. That is what proposals / moves are permissible, when does the game terminate and what is the outcome of a particular playing out.

**Generic game model 2.2**
In the most general case, the game can be viewed as a repeat of the following steps until there are no more eligible moves:
- Does any coupled boy wish to dump his partner? If so, then that boy becoming uncoupled again is an eligible "move", which is realized at that step.
- Does any uncoupled boy wish to make a proposal? If so, then that proposal is an eligible move, which is realized at that step.

When there are no more eligible moves, the resulting matching is the outcome of the game.
|
*Discussion*: There are clearly some problems with this very generic game model above. For example, a boy could endlessly move to become uncoupled and then propose again to the same girl he was coupled with priory. There are other more complex scenarios that can result in an infinite play.

**Assumption 2.2 (no proposals which would clearly be rejected)**
One first simplification which does not restrain generality of the generic game model above is to eliminate from the set of eligible moves those proposals which would be refused (proposals to a girl coupled with someone better than the proposer on her preference list), since, with girls being robotic players under the first few sections, we know these would be rejected and the resulting matching would not be altered.
|
*Discussion*: For simplification of analysis, we sometimes conventionally still say that a boy proposed to a girl who immediately refuses him, as saying that his proposal was rejected. In a strict sense we could just as easily consider that he "skipped" making that proposal and moved on to the next one (if he had any left).

**Assumption 2.3 (no dumping by boys)**
Another simplification we can introduce is that a boy can never spontaneously dump his partner – he needs to be "kicked out" by the girl (who under Assumption 2.1 does so **iff** she receives a better proposal than him).
|
*Discussion*: There are some interesting cases of partial information (either with regard to preference lists or with regard to gameplay of others) where this assumption is too limitative (i.e. it does not model real-life situations). Another potential theoretical issue with this assumption is that a boy might theoretically end up with a worse match (in the "real-life" play) than his anticipated one because some of the members of his coalition did not play "as promised" and therefore nobody "kicked him out" from that "bad" match so he can propose again – while in "real life" he could always dump a partner to make another proposal. But these special cases might be discussed separately. For now, we keep this simplifying assumption.

**Definition 2.4 (naïve / robotic players)**
A boy that plays the game by proposing to the girls in his preference list in decreasing order of preference (should he be refused or later kicked out by his earlier preferences) and never dumps a partner spontaneously is called a naïve player. |

Should all players in the game be naïve players, the resulting game would be the Gale-Shapely algorithm and the resulting matching would be the Gale-Shapely matching. In [1 - College Admissions and the Stability of Marriage, D. Gale and L. S. Shapley, 1962] Gale and Shapley showed that such a matching is Gale-Shapely stable.

**Definition 2.5 (temperature of girls)**
The index of the boy a specific girl $g_j$ is matched to in a particular (final or partial) state of the game, is called **the temperature of girl $g_j$**. |

Notice that for the girls, who are the ones being proposed to, the "quality" of their matching can only increase (or stay the same) over time, as they are proposed to by more and more boys. Therefore:

**Observation 2.6 (non-decreasing temperature of girls)**
The temperature of any girl is non-decreasing throughout the play of the game, under Assumption 2.3 (no dumping by boys). |
*Discussion*: In theory, the temperature of a girl could go down if she is dumped by her then-current partner, thus becoming unmatched. However, for this to happen, the boy that at some point proposed to her needs to have a reason to believe he will be better off by dumping her and (eventually) proposing to some other girl.

**Definition 2.7 (static strategy profiles)**
We say that a player **b** has a static strategy profile $SP\mathbf{b} = <g_{i1}, g_{i2}, …, g_{in}>$ consisting of a permutation of all the girl **iff** he always plays (when uncoupled) to the next girl (to the right) from the one he got kicked-out from or would be refused by and also originally he plays to $g_{i1}$.|
*Discussion***:** Notice that such a player can never propose again to one of his "earlier" girls since, by Observation 2.6 (non-decreasing temperature of girls) that girl would be already out of his reach. Notice also, that a player playing a static strategy profile need not be a naïve player under Definition 2.4: he could just as well not propose the first time to his most preferred girl, for example.
A further noteworthy observation for static strategy profiles is that often-times players may be forced to submit such profiles to some central authority (for example in student-university matching) who then simulates the execution of these strategies to produce a final outcome. The players in such a situation would have no option to play a "dynamic" strategy profile which takes into account "on-the-fly" what happens in the game as other players propose. We shall see that in a range of situations static strategy profiles are sufficiently generic, however not always.

**Assumption 2.8 (all players have static strategy profiles)**
For the first few sections we will assume that all players have static strategy profiles.|
*Discussion*: This assumption was made implicitly by Huang in [10 - Cheating by Men in the Gale-Shapley Stable Matching Algorithm, Chien-Chung Huang, 2006] for example when he considered that strategic play in the marriage problem consisted only of "falsifying preference lists" by the boys who would then be coupled according to Gale-Shapely under these new "falsified" lists. The payoff to each was, of course, the rank of resulting partner under the true preference lists.

**Lemma 2.9 (if all players have static strategy profiles, the order in which they propose is irrelevant)**
Assuming Assumption 2.3 (no dumping by the boys) holds,
The order in which uncoupled boys make proposals, in order, from their respective static strategy profiles is irrelevant to the final resulting matching so long as they will always get a chance to propose when uncoupled.

Essentially we are saying that if we can allow the matching to proceed "step-by-step", allowing at each "step" only one uncoupled boy to make a proposal (to either be accepted or rejected). And that it is irrelevant how we chose the boy of any step: the resulting matching is the same in all cases.

*Proof*:
First, let's brake the simultaneity of proposals by choosing some arbitrary order among proposals that come "at the same time". If the proposals are to the same girl, this clearly poses no problem since the outcome of any arbitrary sequential order will be the same as the simultaneous case: she will keep her best preference among the set. In case there are two simultaneous proposals for different girls, again the outcome of choosing a particular order over them is the same as the outcome of girls choosing simultaneously: each girl will accept the suiter **iff** he beats her then-current match.

Without loss of generality, thus assume that the boys propose in some particular order. For example:

$b_1:f_1, b_2:f_4, b_3:f_2, b_4:f_1, b_1:f_4, b_2:f_2, b_3:f_3, \ldots$
where $b_i:g_j$ signifies "boy $b_i$ proposes to girl $g_j$" at that particular time index in the game.

Notice that for a boy to make a certain proposal at a particular time index he must have either been kicked-out from or rejected by all his "earlier" girls in the static strategy profile.
Notice then that if at a certain time index a boy **b** is at a certain position on his static strategy profile with **g** being the next proposal he would make (if he was the one to "get the microphone" at the particular time index to make the proposal), forever thereafter until **b** finally gets to make his proposal to **g**, **b**'s next play will be at **g**. This follows immediately by the definition 2.7 (static strategy profile). Furthermore, since **b** was uncoupled at that time index, he will for sure get a chance to make his proposal to **g** eventually.

Now we will show that indeed it would make no difference to the final outcome if **b** was to propose to **g** at this earlier time index rather than at the time index he actually proposed at in the initial arbitrary order. Since we chose **b** arbitrarily, as we move from one time index to the next, we can always choose whatever candidate **b** we wish to be the one "who gets the microphone" to propose instead of the one who got it in the arbitrary initial order. Thus, if we prove that "nothing changes" in the final result when we move **b**'s proposal to **g** arbitrarily earlier on (but still to a valid position – so to a time when he is uncoupled and he already has proposed unsuccessfully to all his prior girls in the static strategy profile), then we have shown that any valid arbitrary order produces the same final result.

But indeed, moving **b**'s proposal to **g** earlier on than the moment he actually made it has no net effect. Consider the original segment of play:

prefix $b_t$:$f_t$ middle **b**:**g** suffix, where we have that instead of $b_t$:$f_t$ we would like to have **b**:**g**.

It follows that both $b_t$ and **b** were uncoupled at the time. Then the segment prefix **b**:**g** $b_t$:$f_t$ middle suffix, would be still be a valid play and produce the same result.

Notice in particular that by prefix, middle and suffix we just mean a valid sequence of proposals – the proposals in middle in the first case might have different outcomes than those in middle in the second case. However, they would still remain valid and produce the same end result. We can argue by induction on the time index of a proposal right after **b**:**g**: if the boy say $b_x$ was eligible to propose to say $g_x$ at that time index in the original order, he will also be eligible to propose at that time index (+1 for adding b:g before) to $g_x$. The outcome of the proposal may not be the same of course, but notice that the temperature of girl $g_x$ being proposed to (by $b_x$) is no less that in was in the initial order at that time index (also by induction).
∎

Now, under Assumption 2.3 and 2.8, due to Lemma 2.9 we can introduce an even nicer simplified model of the game which does not restrict generality (produces the same outcome) as the generic model 2.2:

**<u>Simplified Game Model 2.10</u>**
Assume the gameplay is modeled as follows.
A game consists of the following steps:
1. We consider each boy from $b_1$ to $b_n$ in order and for each one we will make the opening play of round **i**, which thus starts with boy $b_i$'s proposal. At the end of each round, all boys from **1** to **i** are matched (temporarily) to some partner.
2. During each round **i** we consider that boys make proposals as such: The (sole) uncoupled boy (from the first **i**) proposes to the next girl in his static strategy profile. This results either in all first **i** boys being coupled (i.e. he proposed to an uncoupled girl) or again some boy (perhaps the same) being uncoupled. The round then continues (with the sole uncoupled boy proposing) until every boy is coupled.

The matching which remained after all n rounds (after boy $b_n$'s round) is considered the final result.
|

We refer to the order in which the boys propose (potentially some more than once) under a set of static strategy profiles as the *canonical order* of proposals.

**Notation 2.11 (play)**
We use the following notation to denote the "play" within any round of the generic game play algorithm described above: …$b_x \rightarrow f_y(b_t)|b_z \rightarrow … \rightarrow f_z$, with the meaning that $b_x$ proposed to $f_y$ and as a result $b_z$ (we can have $b_z = b_x$) was "expelled" from $f_y$ (losing to $b_t$) and went on to propose to some other girl and so on until some boy proposed to $f_z$, who was uncoupled, ending the round. We call each portion separated by **|** an element of the play. Sometimes we omit the parenthesis (like **($b_t$)**) since it is obvious who was the victor in that element by analyzing the next element in the play.
|

**Example 2.12**
Here's an example of the execution of the Gale-Shapely algorithm for example 1.1 in the above notation:
Round b1: b1→g1.
Round b2: b2→g2.
Round b3: b3→g2(b2) | b3→g3.
Round b4: b4→g1(b1) | b4→g2(b2) | b4→g7.
Round b5: b5→g3(b3) | b5→g2(b5) | b2→g4.
Round b6: b6→g6.
Round b7: b7→g4(b2) | b7→g7(b7) | b4→g6(b6) | b4→g5.

The final matching is: b1-g1, b2-g4, b3-g3, b4-g5, b5-g2, b6-g6, b7-g7.
|

Now, what about trading? Given some complete matching, improving upon that would imply that some boy **b1** leaves his current partner **g1** for another, say **g2**, who is left by say **b2** who chooses then **g3**, who was dumped by **b3** and so on, until some boy closes the chain by choosing **g1**.

**Notation 2.13 (trading cycle)**
We use the following notation to denote a switching of partners in a full matching: $f_y|b_x \rightarrow f_z|b_z \rightarrow \ldots \rightarrow f_y$, with the meaning that boy $b_x$ dumps his then-current partner $f_y$ and goes to propose (and be accepted) by girl $f_z$ who was just dumped by $b_z$ and so on, until a boy proposes to the original (now uncoupled) girl $f_y$.

We call such a changing of partners a "trading cycle".
|
*Discussion*: Notice that under Assumption 2.3 (no dumping by the boys), trading cycles cannot be materialized once the matching has been played out. They could however, be agreed on beforehand and the strategy of each player altered accordingly to consider such cycles.

**Example 2.14**
Here's an example of a trading cycle for the Gale-Shapely matching of Example 1.1.

g3|b3→g2|b5→g3.
The new matching is then: b1-g1, b2-g4, b3-g2, b4-g5, b5-g3, b6-g6, b7-g7.
|

Notice that the matching in Example 2.14 is not Gale-Shapely stable, because **b2** is preferred by **g2** over **b3**, her now-current partner and also **b2** prefers **g2** over his current partner, **g4**. Thus **b2** tacit agreement is required for such a matching to materialize.

## 3. MAIN RESULT AND ALGORITHM

It might be useful to refresh the Gale-Shapely algorithm by consulting [1 - College Admissions and the Stability of Marriage, D. Gale and L. S. Shapley, 1962] (for brief overview see also [https://en.wikipedia.org/wiki/Stable_marriage_problem](https://en.wikipedia.org/wiki/Stable_marriage_problem)). The algorithm is a particular execution of the simplified generic game model 2.10, where each uncoupled boy proposes each time to the most preferred girl by him, *he hasn't yet proposed to*, thus being a naïve player (also being a player with a static strategy profile).

The main result of this section is an algorithm which determines a matching which is an optimal result for *all* (i.e. any) boys: in that no other *feasible* coalition could offer any of them a better result, without making someone else (<u>in that hypothetical coalition</u>) *strictly* worse-off.
The main issue (and also difference from the works in [10 - Cheating by Men in the Gale-Shapley Stable Matching Algorithm, Chien-Chung Huang, 2006]) is what we consider to be a *feasible* coalition. We shall substantiate our claim that our assumptions are much more realistic than the implicit ones made in [10].

In this section, Assumption 2.1 (girls are robotic players), Assumption 2.3 (no dumping by the boys), Assumption 2.8 (all players have static strategy profiles) hold. The game model is the Simplified Game Model 2.10.

In this section we shall use the following example problem to illustrate concepts and algorithms:

**Example 3.1**
B = {b1, b2, b3, b4, b5, b6, b7}
G = {g1, g2, g3, g4, g5, g6, g7}

Preference of boys, in descending order (only relevant ones):

| Pb1 | g1 |
|---|---|
| Pb2 | g2, g3, g5, g7 |
| Pb3 | g5, g3 |
| Pb4 | g3, g2, |
| Pb5 | g3, g5, g4 |
| Pb6 | g3, g6 |
| Pb7 | g2, g6, g5 |

Preference of girls, in descending order (only over relevant boys):

| Pg1 | b1 |
|---|---|
| Pg2 | b4, b2, b7 |
| Pg3 | b3, b6, b2, b5, b4 |
| Pg4 | b5 |
| Pg5 | b7, b2, b5, b3 |
| Pg6 | b6, b7 |
| Pg7 | b2 |

I

Note what would an execution of the naïve static strategy profile for each boy (the Gale-Shapely matching) imply for Example 3.1:

**Example 3.2**
Round b1: b1→g1.
Round b2: b2→g2.
Round b3: b3→g5.
Round b4: b4→g3.
Round b5: b5→g3 | b4→g2 | b2→g3 | b5→g5 | b3→g3 | b2→g5 | b5→g4.

Round b6: b6→g3(b3)|b6→g6.
Round b7: b7→g2(b4)|b7→g6(b6);b7→g5|b2→g7.

The final matching would be: b1-g1; b2-g7; b3-g3; b4-g2; b5-g4; b6-g6; b7-g5. |

Notice that the final matching in Example 3.2 is strictly stable (Gale-Shapely stable). Notice however, that it admits a trading cycle: **g2|b4→g3|b3→g5|b7→g2**.

Notice also that under this trading cycle all of **b3**, **b4** and **b7** would get their first choice. Thus, under the works in [10 - Cheating by Men in the Gale-Shapley Stable Matching Algorithm, Chien-Chung Huang, 2006], considering the proposed method described in [11 - Pareto Optimality in House Allocation Problems, David J. Abraham, Katarína Cechlárová, David F. Manlove, Kurt Mehlhorn, 2005] this cycle would surely be part of the final "optimal" coalition. The required "accomplices" in terms of [10] are **b2, b5** and **b6** (all those with legitimate direct veto power, not already in the coalition). But is this "right"? Well, we claim no, for the following reason. In the ensuing hypothetical matching, **b6** could oust **b4** from **g3**. Such a hypothetical possible resulting play would be:
**b6** (dumps **g6**)→**g3**|**b4**→**g2**|**b7**→**g6**.

The final matching would thus be: **b1**-**g1**; **b2**-**g7**; **b3**-**g5**; **b4**-**g2**; **b5**-**g4**; **b6**-**g3**; **b7**-**g6**. Those same-off are in **black**, those better off are in **green**, those worse off are in **red** (if to them there would be no improvement over Gale-Shapely) or in **yellow** (if worse off than in the first case, but still be better off than with Gale-Shapely). This could be a matching directly proposed by some other coalition than in the first case.
The required accomplices for new proposed coalition would be, under the works in [10], just **b2** and **b5**.

Notice that in the latter case we need just a subset of the same accomplices we need in the first case. Furthermore, notice that the only players worse off in the latter than in the former, **b4** and **b7**, do not hold legitimate veto power in the latter. Furthermore, these worse-off players have no recourse in destabilizing the coalition without hurting themselves by risking ending up with a worse match than they would get under the coalition (which is still Gale-Shapely or better for all). Note however that **b7** could "take revenge" by playing at **g5**, but this would entail a worse outcome for himself. This situation will be discussed in the subsequent section. For now, it is sufficient to note that should "the grumpy" **b4** and **b7** simply be naïve players, their cooperation is not required for the coalition (all the rest can simply put their coalition partner – in the latter case – as their first option).

This example, while it does not contradict any theorems in [10] (since in the latter case there are some boys worse off, thus the theorem does not apply), it does contradict the philosophy of what is "the best possible outcome" for each individual boy. For [10], the answer would be the former coalition above for Example 3.1. However, the latter coalition is clearly the feasible one among the two, even if some boys are worse off in it than in the former. This is because those worse off are without power with regard to the proposed outcome, which is something neglected in [10].

Before we provide an algorithm to determine the optimal feasible coalition for some boys, we need to formalize a few notions so as to understand what feasible means for instance. The language used above also employed some definitions detailed below.

**Definition 3.3 (simple coalition)**
A coalition is a tuple <C, $M_c$> where C is a subset of the set of boys B, and $M_c$ is a "promise matching" for the each of the boys in C. The coalition is characterized by the promise of each boy **b** in C to propose first to $M_c$(**b**) and never dump her unless kicked-out (thus when the coalition fails).
|
*Discussion*: In effect, it can be said that boy **b** will propose just to $M_c$(**b**) if the coalition holds.

**Definition 3.4 (direct veto power)**
A player with **direct veto power** over a coalition <C, $M_c$> is a boy **b** such there is a pair (**b1**, $M_c$(**b1**)) in $M_c$ such that $M_c$(**b1**) prefers **b** over **b1**. It is essentially the case that **b** could destroy the promised outcome $M_c$ of the coalition by proposing to a girl which would accept him over the partner to whom she is promised in the coalition.
|

**Definition 3.5 (legitimate direct veto power)**
A player with **legitimate direct veto power** over a coalition <C, $M_c$> is a boy **b** which holds direct veto power, particularly in that he would be preferred over her coalition partner by a girl **g** in $M_c$ who ranks higher in his preference list than $M_c$(**b**). Essentially this is a player that would be accepted by a girl he prefers better than his promised partner.
|
*Discussion*: Note that a player with legitimate direct veto power might not actually get his more preferred girl if he exercises his veto, as the other players react as they are being kicked out, making new proposals (firstly the one he replaces).

A coalition is called **feasible** iff none of the players having direct veto power have an interest to exercise any veto.

There is a thorny issue of determining a motivation for becoming part of a coalition. Or, conversely, determining there is a lack of motivation (lack of interest) to veto a coalition. As we shall see, some boys are "hopeless" – they cannot hope to improve their partner in any coalition. However, they could theoretically still veto some coalitions, at no cost to them (even assuming no suicidal players). All this relates to defining the notion of what is "the best possible partner" for some boy **b**. Clearly a "best possible partner" for any boy **b** could depend also on how the other boys play.

**Definition 3.6 (suicidal play/player)**
We call **suicidal** a play by boy **b** play to girl **g** that is worse in his preference list than the partner Gale-Shapely's algorithm would assign to him, while he still has the opportunity to play to one of his Gale-Shapely partner – or better girls and be accepted.
A player making a suicidal play is called suicidal player.
We call a player that is not suicidal as being **conservative**.
|
*Discussion*: Such a player is called suicidal because, as shown in Lemma 3.7, he could certainly get a better outcome for himself, assuming there are no other suicidal players, simply by being conservative instead of suicidal.

**Lemma 3.7 (Gale-Shapely is worse case outcome with no suicidal players)**
The Gale-Shapely matching is worse-case partner for any boy, assuming no suicidal players.
It is guaranteed that no boy **b** ends up worse off than with his Gale-Shapely partner unless at least some other boy decides to deliberately make himself worse off (at least temporarily) than in the Gale-Shapely matching.

*Proof*:
Assume all players are conservative. Take some arbitrary boy **b**. We will show that **b** can never end up worse-off than in Gale-Shapely.

Since **b** is a conservative player, he first "tries out" girls higher up on his preference list than the Gale-Shapely partner (in some order, not necessarily strictly decreasing order of preference). If he ends up coupled with one of them, then his is no worse off than Gale-Shapely.
Now consider **b**'s proposal to his Gale-Shapely partner **g**. Who can "beat" **b** at girl **g**? Clearly those boys {**b1**,…,**bk**} which rank higher than **b** on **g**'s preference list. However, since **g** is **b**'s Gale-Shapely partner, it follows that none of {**b1**,…,**bk**} ever propose to **g** (otherwise her temperature would go irremediably above **b**) in the course of that algorithm. This means all of {**b1**,…,**bk**} have a better Gale-Shapley partner than **g** (otherwise they would propose to **g** and those she would no longer be **b**'s Gale-Shapely partner).

Now, for **b** to be kicked out of **g**, one of these boys would need to propose to **g**. This is of course theoretically possible (and plays a role in the variant of the game with threat points and ultimatums), but should a player, say **b1**, do this, **b1** would become (temporarily) partnered with a girl worse than with his Gale-Shapely partner and **b** would become unmatched. If **b** does not ever play such that he triggers a chain reaction eventually freeing **b1** from **g** (for example, if **b** plays to an uncoupled girl), **b1** would remain stuck with his less preferred girl, **g**. Why would **b1** ever propose to **g**? He is either suicidal as defined above, sacrificing a certainly better outcome for himself, OR all his partners down until **g** have already been "taken", including thus his Gales-Shapely one. But **b1**'s Gale Shapely partner must have been taken by some **b2**, who again is either suicidal or again had his Gale Shapely partner taken by some **b3** (notice that **b**, **b1**, **b2**, **b3**, … are all distinct since they each ended up with a different partner – namely the prior boy's Gale Shapely partner), and so on. This leads to a direct chain of boys who had their partners "hijacked". Since all boys on this chain are distinct, it follows that the one at the end of the chain has not been hijacked but is suicidal.

Thus, for **b** it is guaranteed that he will get at least his Gale-Shapely partner, unless some player is suicidal, deliberately deciding to make himself worse-off than he could (without requiring any cooperation from anyone other than from all to not be suicidal).
|
*Discussion*: A special case that was easily dismissed was when after having been thrown out by **b1** from his Gale-Shapely partner **g**, the boy **b** makes a play that eventually throws **b1** out of **g**. For example, consider the play **b1→g(b1)|b→g2(b)|b3→g(b3)|b1→**… . Notice that in this case, after irremediably kicking **b** out of **g**, **b1** becomes again free and could just as well propose to his Gale-Shapely partner, with whom he might actually end up coupled (for example if this partner is not **g2**). Thus **b1** has a no-loss threat over **b**: "You will not get **g**!". However, it is required that **b** does not punish **b1** by playing to a girl which would not set in action such a chain reaction freeing **b1** again, otherwise **b1** would himself be worse off. Notice however that **g2**, might as well be **b**'s next preferred girl after **g**. Thus **b** should deviate from being a naïve player if he doesn't get his Gale-Shaply partner if he wants to punish the suicidal **b1**.

### Definition 3.8 (revenge of b4)
Consider the following segment of play
**P = b1→g(b1)|b4→g2(b4)|b3→g(b3)|b1→**…**→g2|b4→**… .
Notice that **b1** kicks **b4** out of **g**. In turn, **b4** indirectly kicks **b1** out of **g**. Furthermore, due to **b1** move's, eventually **b4** also gets kicked out of **g2** (by **b1** directly or potentially by some other player), becoming free to propose to whomever he wants. Essentially, **b4** can take revenge on **b1** for having been kicked out of **g**, with no adverse consequences to himself: he can play to a lesser preferred choice **g2**, knowing that he will eventually also be thrown out of there and thus free to propose to his real next choice after **g**.
We call this situation **revenge of b4**. |

### Assumption 3.9 (no suicidal plays permitted)
For the remainder of this *section*, we assume that players are prohibited from taking suicidal plays.
|
*Discussion*: The assumption above also excludes using threats of suicidal plays as a form of extorting an agreement by some other players for a better coalition matching. Essentially it is as if suicidal plays are not permissible by the rules of the game.

### Definition 3.10 (promises)

We say that a player **b** makes a promise PromNotPlay(**b** | **C**) = {**g1,g2,..,gk**}, if he agrees to bind himself to playing not playing at any of {**g1,g2,..,gk**} so long as condition **C** is met.
While players are free not to make promises, we assume that once they make a promise, it is binding: they are no longer "allowed" to play in a manner which breaks it.

We consider a special case of promise: Coal(**b**, **g**) = Prom(b | $M_c$(**b**)=g) = {$g_i$ | $g_i$ is not **g**} to mean that **b** promises that he will not play to any of his remaining preferences, so long as everybody in the coalition C promises to not play at **g**. Any static strategy profile can easily be adjusted to conform to such a promise, simply by placing $M_c$(**b**) as the first move and then shifting the remaining prefix of moves in some prior profile to the right by one.
|

Now we ask ourselves: what makes a coalition feasible? When can we expect players to be willing to make the coalition promise? We know from Lemma 3.7 and Assumption 3.9 that no one can end up worse off than Gale-Shapely. So clearly no promises by anyone offered less than this outcome will be possible. One intuition is when each player seeks to get "the best possible outcome" for himself. Actually this is a very useful intuition to determine an optimal coalition matching, letting a side for a moment the technical formalization of what "the best possible outcome" means. For example, in the situation discussed around Example 3.1, we noticed two possible coalitions (one giving **b4** his first choice and one giving **b6** his first choice). We noted that both coalitions needed the backing (the coalition promise) of **b2** and **b5**, among possible others. Note that it is conceivable (as is the case with some situations analyzed in subsequent sections) that **b2** or **b5** or both have a strict preference of one of this coalitions over the other, even though they individually get the same outcome in both. Since their cooperation is strictly required for any of these two coalitions to materialize, the first coalition might theoretically end up being the one "played out", if **b6** is convinced for example that **b5** will never agree to the second coalition and always plays at **g5** and thus be certain that his individual outcome in the first coalition is "the best possible outcome". This of course requires that **b6** be content with getting his "best-worse" possible outcome "without a fight" (since **b6** is a required member of the first coalition), giving up a potentially better outcome vetoed by **b5** for no personal gain of his own.

To address such issues, as a starting point, we introduce a further simplifying assumption.

**Assumption 3.10 (self-interest and no taking of sides)**
We consider that each player **b** is interested strictly about optimizing his own individual outcome. As such we require that the sole measure of how "good" a coalition <C, $M_c$> is to him is $M_c$(**b**). Thus, we require that if he "supports" a coalition <C, $M_c$> offering him outcome $M_c$(**b**)=**g** (by making the coalition promise to play at **g** first and never dump her spontaneously) he will support any coalition <C', $M'_c$> also offering $M'_c$(**b**) = **g**.

In essence, this assumption states that a boy will be indifferent how the remaining boys "divide" the remaining girls among themselves, so long as he himself gets the same outcome.
|
*Discussion*: There are interesting cases (with which we will concern ourselves in the following sections) when players have a stake in supporting some other player over yet another. Thus, while they cannot improve their own fate, they might decide to support / veto coalitions such that they favor some other players over the rest. However, as a starting point we disallow such motivations in this section.

### Definition 3.11 (order of preference over coalitions)
Notice that, under Assumption 3.10 a partial order relationship $<_b$ can be defined on the set of possible coalitions $\{<C, M_c>\}$ from the perspective of player **b**'s preference for them. For example, for the two coalitions, let's name them in order C1 and C2, discussed with regard to example 3.1, we can say that C1 $<_{b6}$ C2 and C2 $<_{b4}$ C1.
|
*Discussion*: Under Assumption 3.10, notice that if (not C1$<_b$C2) and (not C2$<_b$C1), then C1 and C2 offer the same outcome to **b** and he is thus indifferent among them (they can be called of equal preference to **b**). Thus we can also define $=_b$ as a relationship and say simply C1 $=_b$ C2. Similarly, we have $\leq_b$.

Everything is nice; players can now compare coalitions to determine which they like better. But how do they find out which (one or several with the same outcome for them) they like best? Again the question boils down to what is an "optimal outcome" for some boy. If there are no coalitions, and all players are naïve the result is bare Gale-Shapely. But the point is to have coalitions. And coalitions require cooperation which entails "settling" for some outcome and not trying to get a better one. But the point of coalitions is to produce some better outcome for the player joining, or, more importantly, not produce a *worse* outcome for him had he not joined.

### Definition 3.12 (infeasible coalitions)
We call a coalition implies some set of player making promises or plays which go against the assumption set of the game model as **infeasible**. |

### Observation 3.13 (coalitions offering worse than Gale-Shapely are infeasible)
Under Assumption 3.9 (no suicidal plays permitted), given our Definition 3.3 (simple coalition) we can notice that any coalition offering a member an outcome *below* his Gale-Shapely matching is infeasible. |
*Discussion*: Not only is such a coalition directly forbidden by the assumptions (axioms) of the game, namely Assumption 3.9, but also it would not make sense anyway (under Assumption 3.9) since by Lemma 3.7 any player who is asked to accept a suicidal outcome could do strictly better by ending up with his Gale-Shapely partner.

### Lemma 3.14 (the hopeless men of a marriage problem instance)

We claim that for any marriage problem, there exists (at least) one hopeless man for which any coalition offering him above his worse-case Gale-Shapely outcome is infeasible.
*Proof*:
Consider all the final men to propose (those who ended up coupled with a then-vacant girl *and were never challenged*) in an execution of the Gale-Shapley's algorithm for the specific problem instance. For Example 3.1, as shown in Example 3.2, they are **b2**, and **b5**. There must be at least one such man. Consider any man **b** from this set, who ends up coupled in Gale-Shapely with **g**.

Assume for the sake of contradiction there would exist some coalition <C, $M_c$>, offering $M_c$(**b**) better than **g**.
Who are **b**'s better preferences? Say {**g1**,…,**gk**}. By the very nature of the Gale-Shapely algorithm, each of these {**g1**,…,**gk**} girls ended up coupled with some boy, say the corresponding one in {**b1**,…,**bk**}. Since by the nature of Gale-Shapley's algorithm, **b** did propose to all of {**g1**,…,**gk**}, (and was rejected or ousted) before settling for **g**, it follows that the temperature of any such girl is above **b** at the end of the execution of the algorithm.
Now say that $M_c$(**b**) = **g1**. Since **g1** prefers **b1** over **b**, it follows that **b1** must himself never propose to **g1**, otherwise, by Observation 2.6 (non-decreasing temperature of girls), **b** will never have a chance. But **g1** is **b1's** Gale-Shapely partner. If <C, $M_c$> were to offer **b1** a worse partner than **g1**, the coalition would be infeasible under Observation 3.13 (coalitions offering worse than Gale-Shapely are infeasible). It thus follows that $M_c$(**b1**) must offer **b1** a match above his Gale-Shapely one (who was just taken by **b**).
Inductively, we can show then that **b1** then takes the Gale-Shapely partner of some **b2**, who in turn takes the Gale-Shapley partner of some **b3** and so on. In essence, we would need to have a trading cycle **g**|**b**→**g1**|**b1**→**g2**|**b2**→…→$g_t$|$b_t$→$g_{t+1}$, where $g_i$ is $b_i$'s Gale-Shapely partner and each $b_i$ leaves $g_i$ voluntarily (he is not ousted by $b_{i-1}$) and moves to the better partner $g_{i+1}$ conveniently vacated by $b_{i+1}$. Notice it is not the case of actual partner-dumping (breaking Assumption 2.3) – since no actual "play" is taken so far, just a simulation of Gale-Shapely – merely a representation of what assigning **g1** to **b** implies. Notice that the final girl on the trading cycle is $g_{t+1}$. She is clearly none of the previous girls {**g1**,…,$g_t$} since she would be "offered" to some prior boy by $M_c$. But since she is necessarily free at this point, and, by induction a better match for $b_t$, it follows that $b_t$ must have proposed to $g_t$ in Gale-Shapely and was eventually ousted. Since she is free now, and is none of {**g1**,…,$g_t$} it follows $g_t$ = **g**, the sole "newly free" girl. But, by our definition of **b**, **b** was never challenged at **g** and never ousted anyone. This means $b_t$ did not even propose to **g** in the Gale-Shapely algorithm. But this is not possible under that algorithm since all boys propose to all better preferred girls than their final match in Gale-Shapely (being naïve players). Thus a contradiction.

It follows that any coalition <C, Mc> offering **b** better than **g** is infeasible. Since **b** was chosen arbitrarily from the set of men who end up with an uncoupled partner unchallenged under Gale-Shapley's algorithm, it follows that all such men are hopeless.

For them, the best possible outcome is the same as the worst possible outcome by Lemma 3.7.
|

Notice that these hopeless men do play a crucial role in most coalitions. Like **b2** and **b5** for Example 1.1. Their cooperation is required for the others to attain a better outcome.

### Definition 3.15 (optimal potential outcome)
We say that player **b** has optimal potential outcome Opt(**b**) **iff** any coalition offering **b** better than Opt(**b**) is infeasible.
|
*Discussion*: The best outcome Opt(**b**) is not necessarily attainable by **b** under any set of assumptions (due to the plays of the other players). However, under the assumptions of this section, we shall see that Opt(**b**) is properly defined for all **b** and furthermore that each player will attain this optimum.

Notice that for the discussion in Example 1.1, both coalitions described are so far feasible under the current assumption model. Therefore, we introduce a new assumption:

### Assumption 3.16 (predilection towards cooperation)
We require that a player **b** supports a coalition that offers him Opt(**b**).
|
*Discussion*: By Assumption 3.10 (self-interest and no taking of sides), player **b** will support any and all such coalitions offering him Opt(**b**). In fact, **b** can materialize this Assumption simply by adhering to the coalition promise that he plays first to Opt(**b**). This way he supports any coalition offering him Opt(**b**). Given that Assumption 2.8 (all players have static strategy profiles) also applies to this section, we can just fix the first element of player **b**'s static strategy profile as Opt(**b**).

Assumptions 3.10 (self-interest and no taking of sides) and 3.16 (predilection towards cooperation) says that essentially a player will strategize only to make himself strictly better off and if he has determined an optimum he can attain he does not care how other players divide the remaining girls so long as he is not hurt.

Notice that both coalitions discussed for Example 1.1 are still feasible under our assumption model (and furthermore both are supported by **b2** and **b5**). It starts to become clear under this model however that the former coalition discussed there (requiring **b6** to cooperate to get a worse outcome for himself) should not be considered "valid", given the existence of the latter one.

**Definition 3.17 (elementary required support set)**
In order for a coalition <C,$M_c$> to hold, and offer outcomes $M_c$, there exists a set $S_0$(<C, $M_c$>) of players who must necessarily support the coalition C by refraining from playing at a girl who would accept them but who is promised to someone else in the coalition. It is thus the set of people with direct veto power over coalition <C,$M_c$>.
|

**Definition 3.18 (current worse case outcome)**
For any player **b** we denote by **Worse(b)** and name **current worse case outcome** the best outcome he could certainly attain given the assumptions of the game model and the current promises in play by other players, irrespective of what other players play (within these assumptions and promises of course).

Under Assumption 2.8 (all players have static strategy profiles), it can be regarded as if, for any potential strategy profile of player **b**, all possible strategy profiles of all the other boys (which are valid under the assumptions and promises of the game) are tried out and the worse outcome is picked for the original strategy profile of **b**. Then the strategy profile with the best such worse case outcome is picked to represent the current worse case outcome.
|

**Definition 3.19 (relative suicidal plays)**
We call **relatively suicidal** a play by a boy **b** to a girl **g1** who is worse off for him than his current worse case outcome Worse(**b**), before having played to Worse(**b**) first.
We call such a player a **relative suicidal player**.
|
*Discussion*: Note that it could be argued that playing to **g** first does not necessarily guarantee player **b** his current worse case current outcome, since the strategy profile which guarantees **g** could entail proposing to some less preferred girl first. However, under the current assumption set we shall see that this is not the case. Furthermore, we can regard this definition as requiring a player to "at least try" to get his guaranteed worse case outcome, before proposing to someone worse.

And finally, our last assumption for this section:

**Assumption 3.20 (no relative suicidal plays)**
Relative suicidal plays are forbidden. There are no relatively suicidal players.
|
*Discussion*: Notice that this assumption explicitly excludes the possibility that some players could use threats of relative suicidal plays to get for themselves outcomes which are far better than for example their then-current worse case outcome under this Assumption 3.20. Remember also the special case about *revenge of b4*, where **b4** hurts other players before ending up same-off as if he didn't. While in the generic sense *revenge of b4* is not self-harmful to **b4**, it is excluded under this Assumption 3.16 (predilection towards cooperation). Thus, right now, players are not allowed to sacrifice the quality of their own outcome in ultimatums.

Note that finally, under Assumption 3.20, the latter (second) coalition in the discussion of Example 3.1 at the start of this section must have the tacit support of **b7** since his worst case outcome (given that **b2** and **b5** cooperate with any coalition) is **g6**, which he attains under this coalition. As such, his "revenge" play at **g5** is excluded as being relatively suicidal. Notice that also **b4** cannot play any lower than **g2** by the same argument. Furthermore, the former (first) coalition becomes infeasible since it would entail **b6** refraining from playing at his thus worse-case outcome of **g3**.

Care should be taken when performing analysis such as the above, for if we were, for example to consider that **b7** gives an "ultimatum" that he will play **g2**,**g5** first, "no matter what", then we will conclude the latter coalition is infeasible and the former materializes. However the sequence **g2**,**g5** is a relatively suicidal play for b7 and thus is excluded under Assumption 3.20.

**Theorem 3.21**
There exists a unique matching $M_c$ for some grand coalition <C, $M_c$>, such that, under Assumptions 3.20 (no relative suicidal plays), 3.16 (predilection towards cooperation) and 3.10 (self-interest and no taking of sides), $M_c(b)$ = Opt(**b**) for all b in B.
Thus, such matching $M_c$ is coalition-stable under these assumptions.
|

Consider the algorithm below:

**Algorithm 3.22**
The following algorithm determines a coalition-stable matching, under Assumptions 3.20, 3.16 and 3.10:

**1**: Let **B**, **G**, **Pb** and **Pg** be the set of boys, set of girls and preferences of boys/girls respectively.
**2**: Set **Result** ← Empty_set.
**3**: While(|**B**|>0) {
**4**: Let (**b**,**g**) be some last pairing produced by running Gale-Shapely(**B**,**G**,**Pb**,**Pg**), namely the element where some boy, **b** proposes last to some free girl **g** and is never challenged there by anyone.
**5**: Set **Result** ← **Result** U {(**b**,**g**)}, thus adding the pair (**b**,**g**) to the final solution, for all such pairs.
**6**: Set **B** = **B** \ {**b**} and **G** = **G** \ {**g**}, thus eliminating both **b** and **g** from the problem, for all such pairs.
**7**: }
**8**: Output **Result**.

*Proof*:

By Lemma 3.7 (Gale-Shapely is worse case outcome with no suicidal plays) we have that Worse(**b**)=**g** for all pairs found at the first iteration of line 4. Also, by Lemma 3.14 (the hopeless men of a marriage problem instance) we have further that Opt(**b**) = **g**.

Thus, at the first iteration the hopeless men of Gale-Shapely get coupled with their best-case-same-as-worse-case partners.

By Assumptions 3.16 (predilection towards cooperation) and 3.10 (self-interest and no taking of sides), all these **b**'s will support any coalition which offers then Opt(**b**). Thus all **b**'s will all make the binding promise to play at Opt(**b**) first. Since Opt(**b**) is also Worse(**b**), this will be their only play under the current assumption set, no matter what other players do (they can never get kicked out from Opt(**b**)=Worse(**b**) since this would entail existence of suicidal players [by Lemma 3.7], which in turn is directly excluded by the more general Assumption 3.20 (no relative suicidal plays)).

Furthermore, by Lemma 3.7 (Gale-Shapely is worse case outcome with no suicidal plays), we have for any boy $b_i$ that the potential match for him by Gale-Shapely in line 4 is Worse($b_i$) under the current (empty) promise set.

By induction on the number of the iteration of lines 4-6, assume that all boys $b_i$ from prior iterations are assigned a guaranteed match of Opt($b_i$) from which they are guaranteed not to be kicked out. The above paragraph argued the validity for the first iteration. Now it remains to be proven that this assumption holds also after the current iteration.

Will any of the boys **b** in the current iteration have an interest in some girl $g_i$ which was offered to some boy $b_i$ at the current or any prior iteration? Clearly not for the girls **g** of the current iteration since the boys they were offered to were never challenged (by **b** in particular) at that girl – so clearly **b** prefers his match **g** to any of them. For prior iterations again we can show inductively that at each step, **b** became coupled by Gale-Shapely in line 4 with a better partner from his perspective than any of the eliminated girls. Furthermore, note (also inductively) that the quality of a potential Gale-Shapely partner determined by line 4 for boy **b** can only improve (or stay the same) from one iteration to another. This follows from the fact that (i) all girls eliminated in line 6 are irrelevant to the remaining boys (induction hypothesis for prior iterations) and (ii) eliminating a boy can never degrade any outcome *for a Gale-Shapely matching*, which is what is produced in line 4 (boys and irrelevant girls are successively eliminated from the problem). Thus no boy **b** will care about any of the eliminated girls $g_i$.

Furthermore, as long as no one "bothers" any of the eliminated boys (by kicking them out of their offered matching), he can also safely ignore them. But none of the remaining boys can bother any of the eliminated ones. Why? Because (again inductively) for each boy **b** still not eliminated, the match produced in line 4 of the current iteration is the then-current Worse(**b**) for him. Thus, by Assumption 3.20 (no relative suicidal plays) his must first make a play at Worse(**b**) before going anywhere lower. But when he makes a play at Worse(**b**) (and of course all other players proposes to their Worse() first before going any lower) he will remain coupled with this girl since, by the very nature of Gale-Shapely all boys who would beat him at Worse(**b**) have a Worse() that is strictly better for them than Worse(**b**), thus will never "get to" propose to her. Therefore, no remaining player **b** can under any (permissible) strategy profile set (of him and the other players) end up proposing lower than Worse(**b**). Thus no remaining boy would ever propose to a girl who is promised to some of the eliminated boys.

By the two arguments above, we have that all remaining boys in the current round essentially do not care about any of the priory eliminated boys and girls. They can never intervene (given their promise also to play at their assigned promised partner first) in any strategy profile permissible under the assumptions and promise set of the game.

Now, for the eliminated boys **b**, can any of them hope to form a coalition where they are strictly better off? Well, since all eliminated boys and girls do not count and can be ignored, notice that **b** is just a hopeless man of a Gale-Shapely round for the remaining boys. By the same arguments used in Lemma 3.14 (the hopeless men of a marriage problem instance), them ever getting a matching better than Worse(**b**) would imply some other remaining boy **b1** gets a matching which is strictly worse than Worse(**b1**). But by Assumption 3.20 (no relative suicidal plays), **b1** cannot support such a coalition (promise to play at his worse-than-Worse(**b1**) partner first). Since **b1** cannot support the coalition, he will always first play to Worse(**b1**) before going any lower. But the moment **b1** played at Worse(**b1**) it is guaranteed that the temperature of this girls is above **b**.

Thus we have shown that also for the eliminated boys **b** of the current round, their promised partners **g**, are both Worse(**b**) and Opt(**b**). Are they guaranteed to keep these partners? Yes, because the induction will hold for all future iterations up until there are no more boys left, terminating the algorithm.

*Complexity analysis*:
The algorithm above consists of at most *n* iterations of Gale Shapley's algorithm plus some O(1) work at each such iteration. The i-th iteration of Gale Shapely takes up to $O((n-i)^2)$ plus some O(1) work. Therefore, the total running time is $O(1^2 + 1 + 2^2 + 1 + \ldots + n^2 + 1) = O(n^3 + n) = O(n^3)$.
The space complexity is dominated by that of Gale-Shapely, namely $O(n^2)$.
|
The proof of correctness of the above algorithm proves Theorem 3.21. Since the algorithm makes no arbitrary choices in producing pairs of the end result, the final coalition stable matching thus produce is the unique one for the instance of the problem.

*Discussion*: Note the crucial part played by Assumption 3.10 (generic cooperation and taking of sides): While it is true that the hopeless man **b** of each iteration is stuck with his partner **g** thus determined, with no hope of any feasible coalition ever improving his matching, he might however have **veto power** as to what coalitions are allowed to form from those which require his cooperation (him playing directly to Opt(**b**) without priory raising the temperature of some girls). So while he has no resolve in making his own fate better (improving his matching), he might have the power to decide which coalitions are allowed to form and which not. Simply put, for a coalition he "likes" he will agree to support it by refraining from proposing to anyone but his final partner, but for all others he might refuse to endorse them and propose to some of his "higher-ups" first. In both situations the hopeless man will be no worse off, no better off, eventually ending up with his final partner. But in case his cooperation is required, he might simply choose to make someone else's faith better off to the detriment of yet another's fate which will be worse off. This kind of preferences are directly excluded by Assumption 3.10: men are assumed to be fully indifferent with regard to what their fellow men's outcomes are, thus allowing the remaining set to "play out" the game as if they (and their partner) were never part of it. This interesting special case of speculating this veto power will be analyzed in a different section.

Here's an example of running the Algorithm 3.22 for Example 3.1.

**Example 3.23**

Iteration 1:
Round b1: b1→g1.
Round b2: b2→g2.
Round b3: b3→g5.
Round b4: b4→g3.
Round b5: b5→g3|b4→g2|b2→g3|b5→g5|b3→g3|b2→g5|b5→g4.
Round b6: b6→g3(b3)|b6→g6.
Round b7: b7→g2(b4)|b7→g6(b6);b7→g5|b2→g7.

The final (unchallenged) propositions are by **b2** to **g7**, **b5** to **g4** and (the irrelevant) **b1** to **g1**. Notice that **b6** is not in this set because he is challenged by **b7** at **g6** and wins. Thus we have Result = {**b1**-**g1**; **b2**-**g7**; **b5**-**g4**}.

Iteration 1:

Round b3: b3→g5.
Round b4: b4→g3.
Round b5: b5→g3|b4→g2
Round b6: b6→g3|b5→g5|b3→g3|b6→g6.
Round b7: b7→g2(b4)|b7→g6(b6);b7→g5|b5→g4.

Iteration 2:

Round b3: b3→g5.
Round b4: b4→g3.

Round b6: b6→g3 | b4→g2.
Round b7: b7→g2(b4) | b7→g6.

The final (unchallenged) propositions are by **b4** to **g2** , **b7** to **g6** and **b3** to **g5**. Thus we have Result = {**b1**-**g1**; **b2**-**g7**; **b5**-**g4**} U {**b3**-**g5**; **b4**-**g2**; **b7**-**g6**}.

Iteration 3:

Round b6: b6→g3.

The final (unchallenged) proposition is by **b6** to **g3**. Thus we have Result = {**b1**-**g1**; **b2**-**g7**; **b5**-**g4**} U {**b3**-**g5**; **b4**-**g2**; **b7**-**g6**} U {**b6**-**g3**}.

Iteration 4:
<there are no more boys remaining>
Thus the final Result = {**b1**-**g1**; **b2**-**g7**; **b3**-**g5**; **b4**-**g2**; **b5**-**g4**; **b6**-**g3**; **b7**-**g6**}, which interestingly enough is the second (latter) coalition discussed for Example 3.1.

|

The method Huang described in [10 - Cheating by Men in the Gale-Shapley Stable Matching Algorithm, Chien-Chung Huang, 2006], which reduces to a single application of Gale-Shapely followed by a single run of the top-trading-cycles implementation in [11 - Pareto Optimality in House Allocation Problems, David J. Abraham, Katarína Cechlárová, David F. Manlove, Kurt Mehlhorn, 2005] takes $O(n^2)$ time. However, we claim that our Algorithm 3.22, whilst running in $O(n^3)$ produces a better quality matching in terms of stability: The discussion in Example 3.1 clearly shows a situation when the matching proposed by Huang's method is not acceptable to some of the required supporters who can, "all things equal" get a strictly better one (excluding relative suicidal play) or in any case either a better or a no worse one by refusing to support the ensuing former (first) coalition discussed, which is the one Huang's method produces.

Note that if Assumption 3.20 (no relative suicidal plays) were to be dropped, a player could gain an advantage in Algorithm 3.22 by submitting a falsified preference list.

**Example 3.24 (lies of b2)**

Say the preference lists are as follows.

Preference of boys, in descending order (only relevant ones):

| | |
|---|---|
| Pb0 | g4, g0 |
| Pb1 | g4, g5 |
| Pb2 | g1, **g3**, **g4** |
| Pb3 | g1, g2, g3 |
| Pb4 | g2, g1 |
| Pb5 | g5, g2, g4, g1 |

Preference of girls, in descending order (only relevant ones):

| | |
|---|---|
| Pg0 | - |
| Pg1 | b4, b5, b3, b2 |
| Pg2 | b3, b4, b5 |
| Pg3 | - |
| Pg4 | b2, b5, b0, b1 |
| Pg5 | b1, b5 |

Assume the players **b2** submitted falsified preference list, switching the order of the bolded preferences. Consider the iterations of Algorithm 3.22.

Iteration 1:
Round b0: b0→g4.
Round b1: b1→g4(b0) | b1→g5.
Round b2: b2→g1.
Round b3: b3→g1 | **b2→g4** | b0→g0.
Round b4: b4→g2.
Round b5:
b5→g5(b1) | b5→g2(b4) | b5→g4(b2) | b5→g1 | b3→g2 | b4→g1 | b3→g3.

The hopeless men are **b3** and **b0**. They remain coupled with **g3** and **g0** respectively.

Iteration 2:

Round b1: b1→g4.
Round b2: b2→g1.

Round b4: b4→g2.
Round b5: b5→g5.

All men are hopeless and get their respective partners.

Iteration 3:
<there are no more boys remaining>
Thus the final Result = {**b0**-**g0**; **b1**-**g4**; **b2**-**g1**; **b3**-**g3**; **b4**-**g2**; **b5**-**g5**;}, where the strategic player **b2** gets his first preference.

Now consider the iterations of Algorithm 3.22 under the unfalsified preference list.

Iteration 1:
Round b0: b0→g4.
Round b1: b1→g4(b0)|b1→g5.
Round b2: b2→g1.
Round b3: b3→g1|b2→g3.
Round b4: b4→g2.
Round b5: b5→g5(b1)|b5→g2(b4)|b5→g4|b0→g0.

The hopeless men are **b0** and **b2**. They remain coupled with **g0** and **g3** respectively.

Iteration 2:

Round b1: b1→g4.

Round b3: b3→g1
Round b4: b4→g2.
Round b5: b5→g5.

The hopeless men are **b1**, **b3**, **b4** and **b5**. They remain coupled with **g4**, **g1**, **g2** and **g5** respectively.

Iteration 3:
<there are no more boys remaining>
Thus the final Result = {**b0**-**g0**; **b1**-**g4**; **b2**-**g3**; **b3**-**g1**; **b4**-**g2**; **b5**-**g5**;}. Notice how in this matching **b2** would get his mere 2nd preference **g3**, instead of **g1**.

*Discussion*: The actions of player **b2** are prohibited under Assumption 3.20 (no relative suicidal plays). However, in case the other boys are not aware of his true preferences, he might get away with it, thus hinting that Algorithm 3.22 might not be suitable for such cases. Note also, that once **b2** played to **g4**, **b5** could play directly to **g3** in his round, in which case **b2** would be a second iteration hopeless man with girl **g4**.

Is there any way can we improve upon the running time of Algorithm 3.22? While intuitively it should be possible, several ideas, like trying to just "update" the Gale-Shapely run from one iteration to another, instead of running from scratch do not lead, after some thought to any faster algorithm. If we will stumble upon a solution on how to compute the matching of Algorithm 3.22 faster, we will publish it in a new paper.

Before we conclude this section, let us introduce a few more definitions and observations which, while not used here directly might prove to be useful for future reference.

Consider a run of Gale-Shapley's algorithm (for example Iteration 1 in Example 3.23 above). Notice that it consists of **n** plays, one for each boy. Each play has an individual element of the form …|$b_x \rightarrow g_y$|… . We will assign to each of these elements a unique numeric ID during the first run of Gale-Shapely.

Like, for example:
**Example 3.25**

Iteration 1:
Round b1: b1→g1(1).
Round b2: b2→g2(2).
Round b3: b3→g5(3).
Round b4: b4→g3(4).
Round b5: b5→g3(5) | b4→g2(6) | b2→g3(7) | b5→g5(8) | b3→g3(9) | b2→g5(10) | b5→g4(11).
Round b6: b6→g3(12)|b6→g6(13).
Round b7: b7→g2(14)|b7→g6(b6)(15);b7→g5(16)|b2→g7(17).
|

Now we will introduce some new concepts for each individual *element*.

**Definition 3.26 (information about elements)**

Notice that the existence of an element …|$b_x \rightarrow g_y$|… implies that $b_x$ first proposed to some better preferred girl in his preference list, let's call her $g_{v-1}$, but was eventually kicked out from or refused by $g_v$ (again, we are running Gale-Shapely here so proposal are made as if from naïve players).
We call the *prior element* to element …|$b_x \rightarrow g_y$|… the element …|$b_x \rightarrow g_{y-1}$|… .
Since $b_x$ eventually departed from $g_v$, it follows that, in the course of the execution of Gale Shapely, some other boy, $b_y$ occupied $g_{v-1}$ at the time $b_x$ was rejected from her (either having been there already or having just proposed). We call this element

…|$b_y \rightarrow g_{y-1}$|… the *direct parent* of element of …|$b_x \rightarrow g_y$|… . Notice that we require the precise element which was still "fulfilled" at the moment of $b_x$ proposal to $g_{v-1}$, not just any of the prior elements which would still beat $b_x$ at $g_{v-1}$. Also we define the relationship Children of an element to be the set of elements for which that element is a <u>direct</u> parent.

We also denote as the *parent set* of an element **x** the set of all elements which entail a proposal to the girl of PriorElement(**x**) and would beat the boy of **x** at that girl.

We also introduce shorthand notations like **x.Girl** and **x.Boy**.

We also call PriorGirl(**x**) for some element **x**, the girl PriorElement(**x**).Girl. Notice that the prior girl for an element never changes (since the prior element also never changes).

We can refer to elements either by their full notation (like …|$b_x \rightarrow g_y$|…) or by the numeric ID we assigned earlier, which will be unique forever.
|

**Example 3.27**
For Example 3.25, we have
PriorElement(b5$\rightarrow$g5(8)) = b5$\rightarrow$g3(5).
ParentSet (b5$\rightarrow$g5(8)) = { b2$\rightarrow$g3(7) ; b3$\rightarrow$g3(9); b6$\rightarrow$g3 (12); }
Or, simply ParentSet(8) = {7, 9, 12}.
Also, we have DirectParent(8) = 7, since then b5 was kicked from his better preference g3 in Gale-Shapely.
Also, note how Children(9) = {10, 12}.
|

**Observation 3.28 (hopeless men are part of an element with no childern)**
A boy **b** of a Gale-Shapely iteration is hopeless **iff** he is part to an element **x** such that ChildSet(x)=Empty_Set.
*Proof*: Notice that an element **x** with no children consists of a proposal by **x.Boy** to girl **x.Girl** who was never challenged there by anyone who lost, not even the initial moment it was made (thus **x.Girl** was uncoupled at that time). Since **x.Boy** is part of the element, it follows **x.Boy** proposed to **x.Girl** and remained coupled with her, never to be challenged or kicked-out from there (since any subsequent or previous play would imply that whoever lost, made a new proposal in some element **y** to which **x** would be the direct parent).
|
Thus, to determine the hopeless men of a certain iteration of Gale-Shapely, it suffices to enumerate the elements which have an empty ChildSet.

Notice that under the relationship DirectParent() the elements (proposals) of Gale-Shapely form a directed tree. For convenience we can introduce some artificial root, say * such that even the first proposals of each boy will have a parent.

In the proof of correctness of Algorithm 3.22 we showed that the matching of each boy **b** in the remaining set can only improve from one iteration to another. Since under Gale-Shapely each boy plays naïve, this simply means that some boys will make *fewer and fewer proposals*, limiting themselves to top something preferences. Notice thus, that never will new proposals appear in a subsequent iteration of Gale-Shapely (after some boys got eliminated) compared to the previous ones. The order in which proposals are made (elements appear into play) could also differ, but the set will never increase.

Also, remember that under Lemma 2.9 (if all players have static strategy profiles, the order in which they propose is irrelevant) we can disregard the order in which the same set of proposals are made, *as long as no boy dumps his partner spontaneously*.

We now pay some thought to the idea of "updating" Gale-Shapley's output under addition/removal of boys.

Addition of some boy is pretty straight forward: we just compute his play based on the prior matching, adding all proposals which he directly or indirectly causes to the tree.

How about removal of boys? Notice that as boys are eliminated some elements of the plays will simply need to be deleted. What happens to the remaining ones? Well, as some boys "move up" in their preference lists, some of those will also need to be deleted as they were *solely* caused (directly or indirectly) by some of the ones removed.

Well, if an element ends up having no possible direct parent, it has no reson d'ete: no reason for being. It thus needs to be deleted also. We need to be careful though that even after some element remained temporarily parentless, there could be some other still "active" element which might very well begin to play the role of parent to that element (i.e. another valid proposal that would beat the respective boy at his prior element's girl). Notice nevertheless that such a potential substitute parent should not come from the subtree of the freshly parentless node (since this would entail cyclic causality).

One might ponder whether an algorithm could be envisioned to dynamically maintain the tree of proposals under deletion of proposal. However, notice that just removing a single proposal (a single element) can cause up to O(**n**) direct parent link changes: for example, this is the case when all n boys propose to the same girl initially and the proposal of the winner is then removed. Notice also that adding a new hopeless man can introduce up to $O(n^2)$ new elements in the Gale-Shapely play (which would need to be deleted when he is removed from the tree).

Consider the following play, for future reference.

### Example 3.29 (inferno of boy b)
Assume b1,b2,…,bn are matched to g1,g2,…,gn before b's round.
Then:

**b→g1|b1→g2|b2→…→g$_{n-1}$|b$_{n-1}$→g1|b→**
**b→g2|b1→g3|b2→…→g$_{n-1}$|b$_{n-2}$→g1|b$_{n-1}$→g2|b→**
**b→g3|b1→g4|b2→…→g$_{n-1}$|b$_{n-3}$→g1|b$_{n-2}$→g2|b$_{n-1}$→g3|b→**
**…**
**b→g$_{n-1}$(b1)|b→g.**

Notice that the play is non-contradictory: no boys proposes to the same girl twice and there exists a feasible preference list of the girls: namely it is the reverse order in which they receive proposals here.

Notice that this play is $O(n^2)$ long and that, in the end, **b** is a hopeless man.

We call this situation *inferno of boy b* (since it degrades the matching of *all* the other boys by n-2 ranks), before ending up proposing to some free girl anyway.
|
Discussion: Notice how tremendous a leverage **b**'s acceptance to play directly to his Gale-Shapely partner has. Him accepting this increases the quality of all the other boys' outcome from second worst to best.

We conclude this section here. Its main result is Algorithm 3.22.

## 4. INTRODUCING THREAT-POINTS AND DYNAMIC STRATEGIES

In Section 3 we made the explicit Assumption 3.20 (no relative suicidal plays) and Assumption 3.9 (no suicidal plays permitted). Making them initially steamed from the idea that no boy is willing to risk getting a worse outcome just to take revenge on or punish some other boy. However, if he would consider taking such potentially self-harmful action, he might be able to just use them as threats. As long as they do not materialize, they are not self-harmful, but, on the contrary, if they produce the desired effect they are beneficial for the one making them.

A realistic case for the situation is that when for example a boy is essentially indifferent among his bottom 90% of the preference list. Thus, as long as he doesn't get a "good" match he doesn't care who he ends up being coupled with. As such, he might very well exploit this indifference by threatening other players (this time at no real potential cost to him) that he will cause them to get worse outcomes than they would if they cooperate with him. This variant of the problem where the preference lists of boys are not strict, but allow indifference, present further complexities and needs to be analyzed separately. However, such a situation could serve as a motivation for a player to engage in blackmail.

But can engaging in blackmail produce better results for a player? We provide an answer by example to this question in the current section.

Thus we ponder what happens when Assumptions 3.20 and 3.9 are dropped and players are allowed to use threats to determine other players to cooperate with them (i.e. "join their coalition").

**<u>Definition 4.1 (ultimative choice)</u>**
We say that a girl Ult(**b**) is such that **b** is satisfied to get Ult(**b**) or any girl ranking higher on his preference list, but he is unsatisfied to get anything below her.
|
*Discussion*: One should not view Ult(**b**) as **b** being indifferent among his top preferences up until Ult(**b**). He might have a strict preference for getting a girl as good as possible. However, if he eventually ends up with Ult(**b**) or better he is called *satisfied*.

Interesting variations arise here as well: does he keep his ultimative preference secret or not? If he does, what does he actually demand of the other players?
Another interesting variation concerns how hard he is willing to hurt himself in order to punish non-cooperators? If he doesn't get his ultimative preference, but could get his next preferred one instead, will he reject this option and hurt himself "all the way" in order to get revenge? How "down" is he willing to go?

**<u>Definition 4.2 (bottom choice)</u>**
We say that a girl Bottom(**b**) is such that a boy **b** has a strict preference for getting Bottom(**b**) or better, rather than getting anything below her.
Namely, he will take whatever strategy so long as it does not land him below Bottom(**b**) in the end.
|
*Discussion*: Bottom(**b**) essentially tells how much **b** is willing to hurt himself in case he doesn't get his ultimative preference. Also, since we dropped Assumption 3.20 (no relative suicidal plays) and Assumption 3.9 (no suicidal plays permitted), **b** might actually play to girls lower than Bottom(**b**), so long as he eventually ends up Bottom(**b**) or better. Notice however that engaging in such play can be dangerous: the moment a boy plays successfully to some girl, under Assumption 2.3 (no dumping by the boys) he might as well end up stuck with her. In fact, the freshly kicked-out player has the option of playing to a particular uncoupled girl, in which case **b** might remain stuck for good.

An interesting case is what would **b**'s motivation be if he doesn't get even Bottom(**b**). What will he pursue then? Utter revenge, "minimize damage" by getting as good a girl as possible under the harsh circumstances, or some mid-way in again punishing some boys but not to the point of hurting himself too much, thus down until some Bottom'(**b**) (then if that fails, down until some Bottom"(**b**) and so on)? Any of these is a valid option – a valid typology of player. Clearly, it is in **b**'s interest to keep his true Bottom(**b**) value(s) secret – this way he can always claim it's the worst possible and thus be able to threaten other players even more.

One further aspect is what happens to the threats **b** makes? In case he doesn't get Ult(**b**) will he actually play out his threats, hurting himself and others? Or will he just settle that his ultimatum was rejected and get the best he can, given that? Clearly, in real life "bluffs" could happen. And sometimes threats could be misread as bluffs and then produce astonishment when they are in fact materialized.

We keep however, the meaning of Definition 3.10 (promises) in that once made, a promise is binding. The player does not have the option of breaking it. Remember that we defined only promises to refrain from playing at some girls, as long as some condition was met. We extend this now.

**<u>Definition 4.3 (complex promises)</u>**
A complex promise P(**b**, T) = S is a promise that if condition T is *ever* met, **b** will limit himself to playing exclusively a strategy from the set S={strategy_profile} and none other. We call condition T the *trigger condition*.
Again, we require that promises are binding. Thus, also, as long as there is an outstanding promise P1 with a potentially still satisfiable trigger condition T1, a player **b** cannot make another promise P2 for some other trigger T2, such that both T1 and T2 are satisfiable simultaneously and the intersection of the strategy profile sets S1 and S2 is empty.
For convenience we allow the definition of set S to be by specifying some set membership function SC, such that SC(strategy_profile_x)=true **iff** strategy_profile_x is in the set S.
|

Under section 3, we had a clear and simple meaning of what a strategy profile was. It was, under Assumption 2.8 (all players have static strategy profiles) a static strategy profile. Now it might be useful to drop this assumption too, in order to allow players to choose whom they want to punish, gradually as the *real-life* (i.e. not any simulation) game is played out (and thus the temperature of some girls is irreversibly raised). We might allow this or not. If the players have to, after negotiating among themselves, submit a static strategy profile to some central authority who then (perhaps assuming they are the boy's true preference list) simulates the execution of the strategy and produces the final outcome, then Assumption 2.8 will hold.
<u>*Warning*</u>: Notice that, given the possibility of strategic play, some central authorities might impose that they automatically provide the coalition-stable matching, or the top-trading-cycles Pareto efficient matching (under the method described in [10 - Cheating by Men in the Gale-Shapley Stable Matching Algorithm, Chien-Chung Huang, 2006]) arguing that it will not make any of the remaining boys worse off than in Gale-Shapely. Strategic play would then imply submitting a static strategy profile that provides the desired outcome under this new, non-Gale-Shapely algorithm. However, we are concerned with how the marriage game would play out in "real-life", with boys proposing to girls under some termination rules, not to the variant where the strategies of propositions prescribed by the boys are no longer respected (as is potentially the case with a non-Gale-Shapely authority).

In order to accommodate a sufficient variety of complex promises, we allow boys to have dynamic strategies. That is, they choose "at each moment of choice" (when they are uncoupled), to what girl they propose next. However, we impose that players to make choices exclusively based on the gameplay, not on other factors such as weather it rained outside or whether they are in a good mood. We can however allow that players introduce randomness in their decision making.

**Definition 4.4 (dynamic pure strategy profiles)**
A dynamic pure strategy profile of some boy **b** is a function
s(**b**) : Game_State → G, associating for each possible state of the game at that moment a girl to whom **b** will propose next.
A Game_State consists of all prior plays all boys made, in the order in which they were made, as well as their outcomes (accepted/rejected). Under Assumption 2.1 (girls are robotic players), the outcomes need not be explicitly specified, as they can be deduced uniquely.
|
*Discussion*: Notice that definition 4.4 excluded the possibility of players deciding probabilistically.

For the moment, we limit our attention exclusively to pure strategies (no random events). But we also define mixed strategies for future reference.

**Definition 4.5 (dynamic mixed strategy profiles)**
A dynamic mixed strategy profile of some boy **b** is a function
s(**b**) : Game_State → Distrib(G), associating for each possible state of the game at that moment a probability distribution over set G of girls, representing the probability with which boy **b** will propose to that given girl. The actual proposal will then be determined at random according to this distribution.
The Game_State information is the same as that of Definition 4.4. Notice that players also have access to prior randomness pertaining to their own choices (since they learn which girl they proposed to actually) and even pertaining to the other players' choices, if their dynamic mixed strategy profiles are known.
|
*Discussion*: Notice that definition 4.5 allowed players to decided randomly at each moment, considering fresh information. This model however is NOT more powerful than if players simply choose a probability distribution over a set of *dynamic* pure strategy profiles. It is however more powerful than choosing a distribution over a set of static strategy profiles.
*Divagation*: If we were to be extremely exigent, we might allow a player to have access to some pool of random numbers and use them anyway he chooses. That way randomness later in the game could be more or less entangled with randomness earlier in the game. But since this complication produces no apparent benefit for the model, we only mention it for philosophical consideration at the moment. Interesting, if maybe absurd cases can be considered if, for example, two players somehow share a source of randomness.

Let us now formalize the game model.

**<u>Dynamic Game Model 4.6</u>**
Assume the gameplay is modeled as follows.
A game consists of the following steps, as long as there exists at least one uncoupled boy:
1. We ask that each uncoupled boy **b** submit his next proposal under his dynamic (pure or mixed) strategy profile, given the current game state. He does not know what other boys decided at the current step.
2. Each girl **g** who received new proposals picks out the most preferred one among that set plus the partner (if any) she already had, forming a temporary couple.
3. The information about what happened in steps 1-2 is provided to all boys.
4. If there are no more uncoupled boys, the game terminates.

The matching which remained after all steps are completed is the final result.
|
*Discussion*: Notice that we require that all uncoupled boys propose simultaneously. That is none of them has the (potential) advantage of knowing what others proposed in that step. He will know only what proposals were made before that step.

Since we allow no "act of consciousness" to interfere in the decision process of a player – i.e. that he decides solely based on the Game_State, not any personal mood or "gut feeling", Definition 4.5 suffices to model the case where any set of complex promises are made beforehand (before beginning of gameplay). In fact, for each feasible Game_State the game might reach, each boy (uncoupled in that situation) will have to produce a next proposal. Allowing him to decide at that moment is the same as asking him to state before-hand what he would do in that situation. Since there can be terribly many Game_States (as we shall discuss) we usually do not require an individual move for each possible Game_State individually, but allow some Game_States to be grouped, producing the same choice (e.g. a naïve strategy can be stated as "for all situations where the boy in question has already proposed to the top x girls in his order of preference, his next proposal is to girl x+1").

The first question we ask is: Can engaging in threat-making help a boy out? It would seem the answer is a big yes. But first, let's adjust our notation to handle the Dynamic Game Model 4.6.

**<u>Notation 4.7 (plays with concurrency)</u>**
We extend Notation 2.11 (play) to handle the new Dynamic Game Model 4.6, where several proposals take place simultaneously and all uncoupled boys have to propose at once. We introduce a small grammar for the play.

We say that a play **P** can be a "conventional" play, under Notation 2.11 (play) with the meaning there stated.

We say that a play **P** can be **P** = **P1**; **P2**; **P3**; …; **Pn**, with the meaning that it consists of the plays **P1**, **P2**, …, **Pn** which started at the same initial moment (the start time index of play **P**). The initial time index for the "root" play is 0.

We introduce the notation **b→Px**, where **b** is a boy and **Px** is a play having the first element include a proposal by boy **b**, as short-hand notation for representing the entirety of play **Px**. So instead of explicitly writing the contents of **Px** there, we just reference it like this.

Also, we introduce the possibility of multiple boys proposing at once, with the notation: **Px** = **b1,b2,…,bk →g(b1)| b→P1**, **b2→P2**, …, **bk→Pk**, to signify that in proposal **Px**, boys **b1,b2,…,bk** proposed to girl **g**, who chose **b1** (dumping her former partner **b**), and then the uncoupled boys **b, b2,…,bk** continued to propose as described in plays **P1**, **P2**,…,**Pk**.
Sometimes, we include the time index of the beginning of a play in parenthesis, for clarity.

Here's an example of using the notation:

P(0) = P1; P2;
P1(0) = b1,b2 →g1(b1)|b2→P3;
P2(0) = b3→g2.
P3(1) = b2→g2|b3→g1|b1→P4
P4(3) = b1→g3.

The root descriptor of the play is **P**.
|

In this section we will use examples without stating explicitly the preference lists of boys and girls (since they may be long), but rather require that they are compatible with some plays we present (under Notation 2.11 (play) or Notation 4.8 (plays with concurrency)). We make the following Observation.

**Observation 4.8 (necessary and sufficient condition for plausibility of plays)**
A play is plausible (there exists at least one set of preference lists by the girls which allows it) **iff** no boy proposes twice to the same girl.
A possible preference list of some girl is the reverse order in which she received the proposals she accepted, with all rejected proposals following down at the end of her list.
|
*Discussion*: Notice that the observation holds even for sets of plays under Notation 4.8.

Now consider the following example:

**<u>Example 4.9 (devilish alliance)</u>**
Let a play **T** consist of the following plays P0 (starting at time index 0), P1 (starting at time index 0), P2 (starting at time index 1), P3 (starting at time index 2), and P4 (starting at time index 14):
<u>T</u> (0) **= P0**; **P1**.
<u>**P0**</u> (0) **= b2→g2; b3→g3; b5→g5; b6→g6; b7→g7; b8→g8; b10→g4.**
<u>**P1**</u> (0) **= b1,b4,b9→g1(b1)|b9→<u>P2</u>; b4→<u>P3</u>;**
<u>**P2**</u> (1) **= b9→g9;**
<u>**P3**</u> (2) **= b4→g2|b2→g1|b1→g3|b3→g9(b9)|b3→g2|b4→…**
      **b4→g5|b5→g3|b1→g6|b6→g5|b4→…**
      **b4→g7|b7→g6|b1→g8|b8→g7|b4→<u>P4</u>.**
<u>**P4**</u> (14) **= b4→g4|b10→g9|b9→g10.**

Let's assign a preference list to boys such that all players except **b4** and **b9** are naïve players in this play P. In case of **b4**, his true second preference is **g10**. In case of **b9** say his true first preference is **g10** and his true second preference is **g9**. Assume **b4** beats **b9** at **g10**.

So what happened in the play?

In P0, all boys proposed to their first choices and were accepted.
In P1, **b9** did not propose to his first preference, **g10**, but instead to **g1**, where he was refused. He did this knowing he would be refused so that he could observe what **b1**'s play at that time index was.
In P2, **b9** still didn't play to his first preference (**g10**) but instead plays to **g9**. He did this to appease **b4** who seeks revenge on **b1** and in turn offers **g9** to **b9**.
Also, in P1, **b1** ousted **b4** from **g1**. However, in P3, **b4**, by not playing at his true 2nd preference (**g10**), took triple revenge on **b1**, causing him to be ousted from not only **g1** but also from his next two choices (**g3** and **g6**), before moving on. Furthermore, he made a lot of other players (**b2**,**b3**,**b5**,**b6**,**b7**,**b8**) worse off.
In P4, **b4** again played to a less preferred choice, **g4**, causing some small havoc for **b10**, *but improving b9's outcome to his first preference*.

We call play T a *threat* by boy b.

Why would such a strange play arise? Simply put: **b4** makes a tempting promise to **b9** that if his instructions are followed to the letter, **b9** will always end up coupled with his first choice, **g10**. Since **b4** beats **b9** at **g10**, **b9** can only hope to get **g10** (**b4's** second choice) if **b4** gets his first choice, OR if **b4** promises to let **b9** have her even if he doesn't end up with his first choice.

Now say that **b4** makes **b9** the proposal to make the binding offer to play according to play T, so long as **b9** promises to cooperate. **b4** could then instruct **b9** to play like this: "For the first round, play to some girl where you would lose for sure [under the assumptions of plausibility of play P, any girl except **g9** and **b10** might do], like **g1**. For the second round, if **b1** played to **g3**, play to **g10** directly. Otherwise, play to girl **g9**." Say player **b9** agrees and both player **b4** and **b9** makes such their promises (which are not just "to each other", but binding).

Now say **b4** makes the following bitter-sweet threat to **b1**: "If you oust me from **g1**, I will exercise triple revenge of **b4**, playing threat **T** and you will get your 4th choice, **g8**. If you leave **g1** to me, I will promise you to get **g3**, your 2nd choice. You need to play directly to **g3**."
If **b1** knows about **b9**'s and **b4**'s already made, binding, promises, he can see that **b4**'s threat-promise is genuine. Say that **b1** accepts, making a binding promise to play to **g3** first.

What happens then? Consider the play Q.
**Q** (0) = **Q0**; **Q1**; **Q2**.
**Q0** (0) = **b2→g2;b5→g5;b6→g6;b7→g7;b8→g8;b10→g4**;
**Q1** (0) = **b4,b9→g1(b4)|b9→Q3**.
**Q2** (0) = **b1,b3→g3(b1)|b3→Q4**.
**Q3** (1) = **b9→g10**.
**Q4** (1) = **b3→g9**.

How do things look now? Everybody gets his first choice, except **b3** and **b1** who gets both get their 2nd choices, **b3** up from his 3rd choice and **b1** up from his 4th choice, compared to the play under threat T.

We call play Q the cooperation alternative to play T.

So everybody should be happy right? **b2**, **b3**, **b5**, **b6**, **b7**, **b8** and **b10** (who were not even required to cooperate beyond a naïve strategy) got better outcomes, **b9** got same **better than Gale-Shapely** (ideal in this case) outcome and even **b1** got significantly better.

So what is the catch? Consider what plays would have occurred had all players been naïve (thus resulting in the Gale-Shapely matching). Of course, naïve over their true preference lists, with no forging.

Consider play G which would occur under Gale-Shapely.

- 40 -

G (0) = G0; G1;
G0 (0) = **b2→g2;b3→g3;b5→g5;b6→g6;b7→g7;b8→g8;b9→g10;b10→g4.**
G1 (0) = **b1,b4→g3(b1)|b4→<u>G2</u>**
G2 (1) = **b4→g10|b9→g9**.

So in the naïve strategy case, **b4** would get his second choice, **b9** also his second choice and all the rest (including **b3** and **b1**) their first choice.

Even applying the coalition stable matching defined under the Assumptions of Section 3, Algorithm 3.22 would produce the same matching.

However, **b4** and **b9** colluding to produce the hell threat **T**, managed to get better outcomes for them both, by employing suicidal and relatively suicidal plays. Notice however, that if we were to modify threat play **T** just by changing P4 to:
**<u>P4</u>** (14) **= b4→g10**.
Then players **b4** and **b9** would both get their Gale-Shapely under threat **T** (in case **b1** refuses to cooperate but **b3** – assuming he could – doesn't exercise some revenge of his own). So they would both be no worse off than in either a stable matching or a coalition stable matching under assumptions of Section 3. And, on the other hand, if their threat **T** holds, then they get the ideal outcome. Why on earth would they not exercise this devilish option?

*Discussion*: We can see how eliminating Assumptions 3.9 (no suicidal plays permitted) and 3.20 (no relative suicidal plays) can dramatically change the prospects for a coalition outcome for some boys – improving them for some, drastically curtailing them for others. Is there any catch? Well, the sole catch so far is the reason we made Assumptions 3.9 and 3.20 – the boys might end up worse off than in Gale-Shapely (or in the coalition stable matching of Algorithm 3.22). The above example was chosen deliberately to allow several plausible preference list of the girls, some allowing say a vengeful **b3** to exercise a revenge of his own in case of threat **T**, throwing **b4** and/or **b9** off from their Gale-Shapely partner, but also some where **b4** and **b9** have absolutely nothing to lose by committing themselves to threat **T**.

As powerful as Example 4.9 is (we may pause for a bit to allow its significance to sink in), notice that it necessarily involved a dynamic strategy profile on part of boy **b9**: he decides his second move based on what **b1** did.

**Observation 4.10**
One question that quickly arises with Example 4.9 is "but couldn't the other boys except **b4** and **b9** improve at least some of their outcomes by playing somehow differently than the naïve strategy? Say knowing that **b4** and **b9** will play the way they do?" The short answer is no, they couldn't. Thus, if **b4** and **b9's** plays are fixed (in terms of a dynamic pure strategy profile), the other players have no means to cooperate to make at least some of them better off without making others (*whose cooperation is needed*) worse.

*Proof*: Under threat **T**, all of **b2,b5,b6,b7,b8,b10** all get their second preference, making just two proposals: one to their respective first preference and one to their second preference. Furthermore, all get kicked out from their first preference as part of a play triggered by a proposal of **b4**. No one on that play would have an interest (in terms of selfish outcome) to play differently since, having already been kicked out from his first choice, the best one could do is the second choice. In case of **b3**, he also gets kicked out from his second choice by **b9**, which is also part of the devilish alliance. So again in case of **b3** the best he could do is getting his third choice, which he does.
|

One question we ask is how to develop the best possible strategy for a player **b** such that, given how he *and other players* would behave, gives **b** the best possible outcome. It would be great if **b** could announce some ultimative strategy, computable based just on the true preference lists of the problem instance such that given the motivations of the other players, he is certain that he will get the best possible outcome for himself (potentially relying on the fact other players will be able to compute this strategy for **b** without **b** needing to announce it, and thus they will adapt given this situation). This problem is far more intellectually challenging than the static strategy profile case with no suicidal play discussed in Section 3. An efficient algorithm to solve it is yet to be developed.

Since any potential benefit from a threat-making strategy is gained only when some parties succumb to the extortion (perhaps considering they are better off giving up a few ranks than "upsetting the madman" and ending up with a much worse outcome), and *alter their own strategies* as a result. Some players on the other hand might find utility in refusing any extortion, even if this would harm their ultimate outcome. Yet others might even try to take revenge (within the game we mean) on a threat-making strategic player.

Notice that this is potentially a game of Ultimatum (see https://en.wikipedia.org/wiki/Chicken_(game) for a brief statement). As such, if some boy **b** fixes his strategy profile (for example by making a promise to play that profile), then the other players, who haven't yet committed to a strategy profile, would have no option than to consider **b** a robotic player for his announced strategy and adapt. Thus, it could be the case that those who "rip off the wheal" first (announce their threat strategy first) can force all other rational players to simply give in to the ultimatum or hurt themselves for sure. Since we concern ourselves with the stable marriage problem and not poker psychology, we will still maintain the requirement that all promises are binding. Thus, a bluff is excluded once a promise is made.
To alleviate the problem of the advantage gained by committing to an Ultimatum first, we allow players to make promises simultaneously. Thus, we augment the Dynamic

Game Model 4.6 to allow any player to make a promise (or a set of promises) at any time during the play, including before the first iteration. Such promises are known to all others, should they wish to consult them. Promise making is considered instantaneous and is permitted to delay the game play (thus we are not troubled by the possibility of some player "filibustering" the game by making an infinite sequence of promises without advancing to a proposal).

We now pay a little attention to what the game play actually consists of, under model Dynamic Game Model 4.6.

**Definition 4.11 (elementary position)**
We call a vector **V** = <$b_{i1}$ or **nill**, $b_{i2}$ or **nill**, …, $b_{in}$ or **nill**> an elementary position in the game, corresponding to each girl in order $g_1$, $g_2$, …, $g_n$ being coupled with the boy specified by the respective position in the vector (or remain uncoupled if that position is **nill**).
It follows immediately that all boys which are not in vector **V** are uncoupled.
We call an elementary position *terminal* if the associated **V** vector contains no **nill**s.
|
*Discussion*: Vector V actually describes the temperature of the girls in play.

Notice that the initial game elementary position is **V** = <**nill**, **nill**, …, **nill**>. Also notice that a "move" at Step 1 of any iteration of the Dynamic Game Model 4.6 corresponds to the selection of one of the (at most) **n** remaining girls a boy hasn't proposed to, for each boy. Such a move would then result, after Step 2 of that iteration, in a new elementary position. We require by convention that boys refrain from proposing to girls who would obviously refuse them, already having a temperature above them.

**Observation 4.12 (Number of terminal positions)**
There are **n!** terminal positions.
*Proof*: Since a boy cannot end up coupled with two girls, and all boys are coupled, the resulting terminal vector **V** must be a permutation of the boys (as we assumed without loss of generality an equal number of boys and girls).
|

**Observation 4.13 (Total number elementary positions)**
There are can be no more than $(n+1)^n$ elementary positions in total.
*Proof*: The maximal bound follows immediately from the definition of the position vector. Each element in the **n** element vector **V** can have at most **n+1** distinct values (**n** boys plus **nill**).
|

**Observation 4.14 (The game graph is a DAG)**
The game graph is a directed acyclic graph with one root node and **n!** terminal nodes. We consider that edges are labeled with one element from the potential moves by all uncoupled boys at that particular position. The moves are not directly restricted by prior gameplay other than the then-current elementary position of the game.
*Proof*: Each time a move is made in the game, the temperature of at least some girl **g** increases (potentially from **nill**), since we exclude futile proposals by boys. The incoming edges for an elementary position (except the root one) are from a position where **g** had a lower temperature. Thus, for a cycle to complete, one of those positions would have to be reached again. However, by Observation 2.6 (non-decreasing temperature of girls), still valid under Dynamic Game Model 4.6, girl **g**'s temperature cannot decrease throughout the remainder of the game. Thus, none of the position's source (inbound) nodes will be reachable. Thus no cycle is possible.
∎

*Discussion*: A gameplay is actually a path in this DAG from the root node (**V**=<**nill**,**nill**,…,**nill**>) to some terminal node. However, note that under Dynamic Game Model 4.6, a player is allowed to decide based not just on the current *elementary* position (temperatures of girls) but also on the history of the game. Thus, to be complete, a position in the game would consist of an elementary position PLUS the path the game took to reach it (the edges, in order).

**Definition 4.15 (Game_State)**
We say that a that a Game_State S is a pair <**V**, **H**>, where **V** is an elementary position and H is an ordered list of edges which were followed for the game to progress from the initial elementary position <**nill**,**nill**,…,**nill**> to elementary position **V**.
∎

**Observation 4.16 (Size of history)**
The maximal size of the history H of some Game_State S is $n^2$ ordered proposals.
*Proof*: Note that the Game_State's history is in fact a chronological arrangement of proposals, some of them grouped together at the same chronological time (thus forming an edge). Notice that for a given chronological list of proposals, it is immediate which proposals form an edge (since all uncoupled boys after the previous edge must propose in the current edge and none other than them). By Observation 2.6 (non-decreasing temperature of girls) and Assumption 2.3 (no dumping by the boys), there can be at no more than $n^2$ proposals in a game play (each girl receiving a proposal from each boy).
∎

**Observation 4.17 (Number of distinct histories)**
The worst case number of game histories for a problem instance is between **(n-1)$^n$** and **(n$^2$)!**.
*Proof*: By the above argument it quickly follows there can be no more than **(n$^2$)!** game histories (at most one for each permutation of the at most **n$^2$** proposals, with some permutations not representing a possible game play). Since each game history also uniquely determines the resulting elementary position, we can say there are no more than **(n$^2$)!** Game_States. Under this estimate, for our Example 4.9, there could be more up to more than $10^{157}$ Game_States. But this maximum may not necessarily be reachable.

Consider Example 3.29 (inferno of boy b). Under the induced preference list of girls, each time a boy is uncoupled (up to **n** times), he might as well propose to *any* of the girls (except the last); the final outcome would be the same – he only shortens the game history a bit: with the number of "skipped" girls in his preference list (who may end up higher than him in temperature next time he is uncoupled) and the number of potential proposals by all boys ranking below him at that respective girl, who have not yet proposed. We can say that a boy can decide to shorten the history each time he proposes by a number between 0 and at most some value below 2**n**. Even if each boy was to decide initially that he would only once skip a number of proposals (between 0 and **n-2** for all boys except **b** whose range is between 0 and **n-1**), most combinations of such decision would result different histories (as players decide to skip, the remaining players may have their options to skip become more and more limited). There are **(n/n-1)**$^*$**(n-1)**$^n$ possibilities. For 11 boys this value would be $>10^{11}$. Still way too many. And this is just considering a single decision to "skip ahead". In the situation of Example 3.29 (inferno of boy b), a boy could make successive decisions to "skip ahead" some of his proposals.
|

We can quickly note that solving the game by solving each potential Game_State is not feasible, even though such an approach can lead to an algorithm for the optimal solution, if one exists: Since the game graph is a DAG, one can color nodes, starting at the leaves (terminal nodes) based on weather that said node is a "winning", "losing", acceptable or inacceptable (below Bottom(**b**)) position for any player **b**. Save coordination issues, a player could then decide at each position for a branch that would leave him best-off. Since such an algorithm would run in super-exponential time, we do not proceed further in this direction. However, even though the number of Game_States makes solving the Game at each position unfeasible, we can still use this DAG modeling of the game to reason about how players strategize.

An Ultimatum is useful only when it is accepted. One cannot develop an adequate model for developing an optimal strategy without knowing when ultimatums are accepted and when they are rejected. By definition of Bottom(**b**), we require that each player accepts an Ultimatum which would – when rejected – leave him bellow Bottom(**b**), but when accepted allow him to get Bottom(**b**) or better.
Again, we have not clarified what happens when a boy gets below Bottom(**b**) [no matter what]: that is, given the strategies of the other players, he cannot possibly end up Bottom(**b**) or better. We allow his play to be arbitrary in that circumstance. In particular, the natural expectation is that he will try to get other boys below their Bottom() outcome, in the hopes of deterring them from the plays that got him (**b**) below Bottom(**b**). Note however, that perfectly valid problem instances exist where some players always end up below their Bottom() and they do not hold any direct veto power over the outcomes of some coalition, leaving them powerless.

For the moment we concern ourselves with when **b** can get Bottom(**b**) or better.

### Assumption 4.18 (implicit promises)

If all potential feasible paths from a given Game_State down the game DAG end up in some terminal position where a boy **b** ends up worse than Bottom(**b**) we say that such a Game_State is *losing* for boy **b**. A path is feasible if given the promises and assumptions of the game model, all boys are permitted at all times to make the proposals resulting in that play path.

Every boy **b** makes a mandatory initial promise that whenever he is in a non-losing Game_State he will never make a play that, given the promises and assumptions of the game model, would *certainly* lead to one of the losing Game_States for him.

Note that the initial "wrong move" that **b** might ponder to make need not be the proposal part of the move immediately leading to the losing Game_State, but one further "above" it in the game DAG.

Furthermore, we require that any player who ends up below his Bottom(**b**) will take such a strategy as to deter the prior play by some other boy or boys, which caused him to get below his Bottom(**b**), by also getting them below their Bottom(). Assuming, of course, such a play exists.
|
*Discussion*: Note that the above assumption is quite limitative. First of all, each boy **b** implicitly undertakes that he will NOT be willing to hurt himself below Bottom(**b**) in order to deter others from a particular game path (which might result in him getting far above Bottom(**b**)). This was, after all, the definition of Bottom(**b**). Secondly, player **b** might find himself in a losing Game_State at the beginning of the game – before the initial proposals. In such a situation we say the game instance is **invalidated**. Also note that this assumption implies a promise by **b** to refrain only from moves which are *certainly* losing for him. Since at any given iteration of the Dynamic Game Model 4.6 several boys decide at once, if their decisions are not fixed by promises, a certain proposal by boy **b** might or might not lead to a losing position for him, depending on what the other boys simultaneously chose. Thus, under Dynamic Game Model 4.6, the marriage game is also a game of coordination. We call a game instance where the game play could result in such a situation, where a boy is faced with a situation where he may play resulting in a losing Game_State for him, as **invalidateable**.

Note that an invalidateable game instance might no longer be invalidateable if the players other than **b** (from the above discussion) change their promise set so that, should the game reach that strange situation where **b** can make a play that may or may not be losing, they all play such that **b**'s play is losing. Also note that such a promise set may not always exist, since **b** might have more than one valid potential proposal which might lead or not to a losing Game_State for him and thus the other players may not know which one **b** actually makes (under a dynamic mixed strategy profile for example). **b** in turn might want to decide based on what the other boys actually play, hence a cyclic dependence. This is the key trait of a synchronization game.

We shall seek to avoid both invalidated and invalidateable game instances.

**Assumption 4.19 (further implicit promises)**
If all potential paths from a given Game_State down the game DAG end up in some terminal position where a boy **b** ends up worse than Ult(**b**) we say that such a Game_State is *unsatisfactory* for boy **b**.

Every boy **b** makes a mandatory initial promise that, he will never make a move *certainly* leading to an unsatisfactory Game_State, so long as he has another move which *certainly* leads him to a satisfactory position – given the promises and assumptions of the game model.

Note again that the initial "wrong move" that **b** might ponder to make need not be the proposal part of the move immediately leading to the unsatisfactory Game_State, but one further "above" it in the game DAG.
|
*Discussion*: Essentially unlike Assumption 4.17 where each boy promised never to make a play certainly leading him to a losing position, here we require that each boy promises never to play to certainly an unsatisfactory position, *so long as he has a certain strategy to attain a satisfactory outcome*. Note that this is much weaker than the former assumption (for Bottom(**b**)) since a boy is allowed to take strategies which would certainly lead him below Ult(**b**) if he has no guaranteed one leading him at or above. Furthermore, if he ends up below Ult(**b**), the game path does not get invalidated (as in the case of Assumption 4.17), but the promise no longer has any effect (and **b** can potentially take revenge including by hurting himself down until Bottom(**b**)).

Note that both Assumption 4.17 and 4.18 together do necessarily limit the strategies available to any single player to just one.
Since these assumptions imply promise making (and thus a potential limitations of gameplay), a totally "free" player would be inclined to choose Ult(**b**) as his first preference and Bottom(**b**) as his last preference. That way, he either gets his first choice, or he is free to play however. But if he cannot get his first preference no matter what, then what? Logic of self-interest would dictate he would lower Ult(**b**) to his second preference and see if he can attain that. But really?

**Definition 4.20 (augmented problem instance)**
We say that an augmented problem instance for the marriage problem includes, alongside the boys' and girls' preference lists, the Bottom(**b**) and Ult(**b**) specifications for all boys **b**.
|
*Discussion*: Notice that in real-life situations, only Bottom(**b**) may actually be "hard-fixed" apriori. A player may be flexible with his Ult(**b**) choice. But then again maybe not.

**Example 4.21 (what a standoff)**

Note that for Example 4.9 (devilish alliance), if both Ult(**b4**) = **g1** and Ult(**b1**) = **g1**, then they cannot both get their sole ultimative choice. Furthermore, if any of them lowers his choice Ult() one notch (**b4** to Ult(**b4**) = **g10**; or **b1** to Ult(**b1**) = **g3**), then both options are feasible. Furthermore, they both get their second choice if they both lower it. So who should lower it then?... In Example 4.9, in case of non-cooperation **b1** would get his 4th choice rather than his 2nd one if he cooperated (and **b4** potentially say his 3rd at **g4**). But the preferences could be adjusted so that, in case of non-cooperation by **b4** (if **b1** is kicked out of **g1**), **b1** might play to **g4** (assume successfully) and **b4** would then end by stuck with **g2** (maybe his 4th or 5th or 6th choice – say his 5th choice). This is a typical standoff. Who will back off? **b1** or **b4**? If neither of them backs off, then they both get worse outcomes (lose-lose situation). But it may very well happen! We call such a situation *a standoff*.

In fact, this standoff situation is a distinct coordination game – an Ultimatum game called Game of Chicken (see [https://en.wikipedia.org/wiki/Chicken_(game)](https://en.wikipedia.org/wiki/Chicken_(game)) for a brief description). There is nothing in the data of even an augmented marriage problem instance which might determine how such a standoff is resolved, in the general case (for example in case for example both Bottom(**b4**) and Bottom(**b1**) are set to their respective last preference, in which case no player is forced by Assumption 4.18 (implicit promises) to avoid that play).

Another noteworthy aspect is that if dynamic mixed strategy profiles are allowed, the players might even agree to "flip the coin" over the decision to pursue **g1** or not. Note how using a shared coin might help them synchronize (so they don't both get to propose to **g1**), thus making the *Divagation* of Definition 4.5 (dynamic mixed strategy profiles) interesting. In this particular case (where they both propose to the same girl), *if they trust each other*, there is no mandatory need for a shared coin. However, in other cases such a device might prove useful (for example if they commit to one of two strategies by first proposing to some other girl **g12** or **g13** and **g14** or **g15** respectively), even if the players trust each other (in making a truly random choice).

Given the potential for standoffs, even an augmented problem instance may not have a definable optimum for some players. If the augmented problem instance contains standoffs, without being given some apriori method of resolving them (for example an ordering of the boys based on their willingness to back off in a stand-off), what one could hope for is to determine that there is a *reachable* standoff and leave it at that. Note however, that the mere existence of standoffs need to threaten the optimal outcome of some boys not directly or indirectly involved with the standoff. If Example 4.9 (devilish alliance) were to be augmented with some other boys and girls, with the newly added boys having girls **g1**-**g10** as their worst preferences, they couldn't care less how **b4** and **b1** decide their standoff – they would be unaffected (so long of course as none of the first boys can and decides to involve them with the cost of making himself yet even worse off).

We note that in a non-augmented problem instance there can always be stand-offs: A player **b** who plays to **g**, below his Gale-Shapely outcome, can end up with that match if no other boy proposes to **g**. In fact, if the freshly kicked-out boy can propose to the girl who would end up uncoupled under a Gale-Shapely for the remaining set of players. That way, if no other players except **b** are threat-making, this player will not contribute to **b** being kicked out of **g**. Even if there are other threat-makers, if all blackmailed boys play according to this strategy, all threat-makers will end up worse-off then their Gale-Shapely. Thus, there is an inherent stand-off between any threat-yielding player and the rest: if his ultimatum is rejected, he can be made to end up worse off than if he drops his threat strategy.

Using a maximum matching algorithm, one can determine if an augmented problem instance admits a solution *satisfactory* to all players (when they all get above their **Ult**()). This can be verified using a bipartite matching algorithm: each boy can be matched with any of his Ult(**b**) or better preferences. The augmented problem instance admits a *satisfactory* outcome for all players **iff** there exists a perfect matching (matching all **n** boys). We refer to the Ford-Fulkerson method [14 - The Ford-Fulkerson method, TH Cormen, CE Leiserson, RL Rivest, C Stein, 2001] for solving maximum bipartite matching. Better methods may exist. Since the maximum size of the matching (and thus maximal flow in [14]) is **n**, and each boy has at most **n** preferences, the running time would be $O(n * n^2) = O(n^3)$.

Note however that such an approach only gives a hint as to what the outcome of optimal play may be. Under optimal play, it may well be that some players can get satisfactory outcomes no matter what the other players do. For Example 4.9 (devilish alliance), if Ult(**b1**)=**g1**, but all other players have no Ult() [formally their Ult() is their least preferred girl], even if **b1** does not get **g1** under the threat play, all the others get at or above their Ult(). In fact, the preferences of the girls in Example 4.9 could be tweaked so that the Ult() of all players except **b1** and **b4** are their actual outcomes under the threat play (their second preferences most of the time); this way even if **b1** gets unsatisfactory outcome, they still all get satisfactory ones.

Furthermore, a problem instance can admit several solutions satisfactory to all players. Some in which some group is better off and other in which part of this group is worse off. So which one should be chosen? As with the case with the approach in [10 - Cheating by Men in the Gale-Shapley Stable Matching Algorithm, Chien-Chung Huang, 2006], choosing any and improving it by the top-trading-cycles method improperly ignores the power players have to object to being part of coalitions, as discussed in Section 3. Worse even, it ignores the potential of threat-making to deter an outcome.

**<u>Definition 4.22 (undeterable coalitions)</u>**

We say that a coalition Coal = <**C**, **M_c**> is *undeterable* if no player or coalition of players outside Coal.**C** can force some player in **C** to end up matched with a girl outside **M_c** or with a girl in **M_c** who is below his Bottom().

Note that a player **b** may be forced from **M_c**(**b**) under the definition, but not from all girls in **M_c** who are above his Bottom(**b**).

We also say that a coalition Coal is *undeterable* (or holds control over a set of girls) *with regard to a set S of external players* if definition 4.22 (undeterable coalitions) holds when the set of external players (**B**\Coal.**C**) is limited to just **S**.

|

*Discussion*: We say that the set of players **C** *holds control* over the set of girls in **M_c**.

### Lemma 4.23 (condition for undeterability of coalitions)

A coalition Coal = <**C**, **M_c**> is *undeterable* if there exists a "worst case" perfect matching **M_w**, *reachable* from **M_c**, defined as follows: Each boy **b** in **C** gets an edge to all girls **g** in **M_c**, which satisfy (i) **g** is at or above Bottom(**b**); (ii) there are no external players (in **B**\Coal.**C**) holding direct veto power over **b** at **g**.

The converse is true for non-augmented problem instances (thus dropping condition (i)).

We say that the matching **M_w** is *reachable* from **M_c** iff should any boy **b** be kicked out from **M_c**(**b**), he can trigger a play, which comprises only girls in **M_c** and culminates with **M_w**$^{-1}$(**M_c**(**b**)) at **M_c**(**b**) and the matching **M_w** is still reachable from the new matching (or coincides with it).

*Proof*: '➔' Suppose some external player in **B**\Coal.**C** played successfully at some girl in **M_c**, kicking out boy **b**. Then **b** can generate a play the play which places **M_w**$^{-1}$(**M_c**(**b**)) at **M_c**(**b**) and still preserves **M_w**'s reachability, consisting of girls only in **M_c**. Such a play preserves the conditions that all girls in **M_c** are occupied by coalition boys. Ultimately there will be no more direct vetoes (for example the matching becomes **M_w**).

'⬅': If there exists no such perfect matching **M_w**, then the external players have an easy strategy: So long as there exists an uncoupled external boy holding direct veto power over some girl in **M_c**, he exercises it (he makes that proposal). All such direct vetoes are exercised. What can be the end result? Either some external boy ends up occupying a girl in **M_c**, OR there are no more external boys holding veto (since all external boys exercise their vetoes first, they cannot end up coupled outside **M_c** so long as they still have a direct veto power). In the second case, the resulting matching of boys in Coal.**C** to girls in **M_c** would meet the conditions of the perfect matching **M_w**. Thus, the first case in which the coalition outcome was successfully vetoed must be true.

|

*Discussion*: Note that the converse statement is not necessarily true for an augmented problem instance. There can be undeterable coalitions for an augmented problem instance which do not satisfy the conditions of Lemma 4.23, because some boys might be restricted from taking certain plays by existence of their Bottom(). Also, although a coalition might be deterable, the players which hold direct veto power over it might have no interest to exercise it (for example, they may get their first choice on a girl outside those in the coalition matching). However, the only guaranteed outcome is the worst case $M_w$ matching in a coalition which is undeterable.

Note that Lemma 4.23 also provides an $O(n^3)$ worst case method to determine if any fixed set players *holds control* for sure over any fixed set of girls, under the criteria of the Lemma: A maximum matching algorithm is run on a graph instance constructed as described in Lemma 4.23, with the initial temperature of all the girls being **nill** (they are unmatched). The total time to construct such a graph can be as low as $O(n^2)$ (it is $O(n^3)$ naively) and running the matching algorithm takes an extra $O(|C|^3)$ worst case, for a total running time of no worse than $O(n^3)$.

When a coalition has control over a set of girls, the only reason some other external player(s) would play at the one of the coalition girls is to degrade the outcome of some players in the coalition, maybe down to some $M_w$. But if that worst case outcome is acceptable to the boys in the coalition, then such a vengeful action will produce no results. However, it may also not entail any cost: the vetoing player(s) will be uncoupled in the end again. Then again, should some boys in the coalition choose to punish the vetoing player(s), then he (or them) might actually end up worse-off than if he had not vetoed.

**Definition 4.24 (external stability of matchings)**
We say that for a coalition Coal = <**C**, $M_c$>, the coalition matching $M_c$ is *externally stable* **iff** there are no external players holding direct veto power over it.

Again we expand the terminology to include external stability *with regard to a set S of external players*.
|
*Discussion*: Clearly, the existence of an externally stable matching implies the coalition *holds control* over the girls in $M_c$. However, not all matchings over these girls to the coalition boys may be externally-stable, or result in an undeterable coalition.

Note also that a coalition can safely play directly to any of its externally stable matchings. No external players would be able to alter the outcome. However, some matchings which are not externally stable could result from normal play (for example when the veto holders are best preferred by their own first preference which lies outside the girls divided by the coalition).

Note that a coalition having control over a certain set of girls, might choose to include new players which, in exchange for their contribution to the overall control over the new set of girls, are promised some girl from the initial set (which they could not get without cooperation of players in the coalition).
For Example 4.9 (devilish alliance), note that the boys **B**\{**b1**} have control (under some set of girls' preferences compatible with the plays) over the girls **G**\{**g8**}. As such, if such a coalition forms, **b1** cannot get any of these girls and has to settle for **g8**.

However, he still can wreak some havoc first, by vetoing **b4** from **g1**, etc., as illustrated in the Example. However, should **b1** not be satisfied with **g8** (for an augmented problem instance, maybe **g8** is below his Bottom(**b1**)), the coalition could propose to him by offering him **g2**, in exchange for cooperation. In this manner, the cooperative outcome of Example 4.9 (devilish alliance) can ensue. Furthermore, one can say that that coalition **B\{b1}** has a threat over **b1** above **g8** (they can prevent **b1** from getting better than **g8**). Note that although {**b4**} might have control both of {**g4**} and {**g10**}, he cannot alone have control over {**g4**, **g10**}. Thus, while the coalition **B\{b4}** does not have control over any set (since any set of appropriate size encompasses either **g4** or **g10**), it still can guarantee that the outcomes will be from a certain set. Furthermore, the coalition **B\{b4,b9,b10}** has control over {**g1,g2,g3,g5,g6,g7,g8**}. However, the externally stable matching **B\{b4,b9,b10}** can offer is either the same as that by **B\{b1}** or worse for some players. Which coalition can prevail to convince the rest to cooperate? The first must convince **b1**; the second must convince **b4**. Hence a standoff situation.

We now examine an approach which seeks to address the situation around standoffs.

### Definition 4.25 (agreements)
A special kind of promise, is "a promise over a promise". That is, player **b** promises **P** with the condition C = "if some other specific player **b2** (or set of players) makes/doesn't make another specific promise **P2**", and the content "I will then also make (a distinct) promise **P1**.".

Such a promise **P** is essentially a binding offer (or threat) of agreement: **b** binds himself that if **b2** promises **P2** he will then promise **P1**. Since promises are binding, there can be no cheating in an agreement. If **b2** makes promise **P2**, the automatically **b**'s promise **P1** will be in effect.

Futhermore, **b** could also promise that if **b2** offers the converse promise (to bind himself to promising **P2** if **b** promises **P1**) then he will make the promise **P1**. This way agreements could be reached simultaneously.
|
*Discussion*: We introduce agreements to allow the posibility that players can force themselves to respect joint strategies even when their threats are rejected. Since "keeping a promise" is always possible by any player voluntarily, there is not always a need for some central authority or "law of nature" to prohibit breaking of promises. It could be that players adequatelly describe their mental states which will not change in that regard. Notice again that making promises is not required. Some players may make no promises at all.

One can attempt to determine some theoretical optimal potential outcome for some fixed boy **b**. Consider the algorithmic approach below.

### Algorithm 4.26 (expanding wrath approach)

The following algorithm seeks to determine if there exists a possible resolution to standoff situations where boy **b** gets girl **g** or better. Initially the routine below is called with Coal = Empty_Set.

**0**: Initialize Lower[**b**]←**nill** for all **b** in **B**.
**1**: bool IsOutcomeFeasible(**B**, **G**, **Pb**, **Pg**, **b, g,** Coal) {
**2**: Set Coal.**C** ← Coal.**C** U {**b**} and Lower[**b**] ← **g**.
**3**: Does there exist a perfect matching of each boy **b2** in Coal.**C** to a girl **g2** which at or better preferred than his respective Lower[**b2**]? If not, return **false**.
**4**: Does there exist a perfect matching of each boy **b2** in Coal.**C** to a girl **g2** which at or better preferred than his respective Lower[**b2**], which is *externally stable*? If yes, return **true**.
**5**: For each boy **b1** in **B**\Coal.**C** { //Try to expand the coalition by this boy
**6**: Let **Sx** be the maximal set of top preferences of boy **b1** (i.e. his top **x** preferences), such that Coal.**C** has control over this set **Sx** with regard to {**b1**}. //This is the set from which **b1** can be expelled by the coalition.
**7**: Let **g1** be the least preferred girl by **b1** in set **Sx**.
**8**: If(IsOutcomeFeasible(**B**, **G**, **Pb**, **Pg**, **b1, g1,** Coal)) return **true**;
**9**: Lower[**b1**] ← **nill**.
**10**: }// line **5**.
**11**: return **false**;
**12**:} //line **1**.

*Discussion*: The algorithm essentially considers all permutations under which a coalition can grow from just <{**b**}, {**b** – some girl at or above **g**}> to an undeterable one, such that, at each expansion, the new boy joining the coalition is offered something he certainly could not get if the coalition thus-far was hostile to him. The algorithm thus can be used to determine the proper order in which boys could be made "bitter-sweet" offers before the beginning of the game, to tempt them to join the coalition. In case some boy along this feasible path refuses, he can be punished by the boys in the coalition. However, if he cooperates and all future boys cooperate he will get a better matching. Since the first non-cooperator will be the one who will be surely worse-off, no one has the interest to be the first to refuse cooperation. Inductively, all boys (including future ones) will cooperate, if the algorithm returns true. This, of course, assumes that none reject the "bitter-sweet" Ultimatum of the coalition. For this, all ultimatums have to be credible: should an ultimatum be rejected, the boys in the coalition must be prepared to occupy all girls in the respective **Sx** set. For an augmented problem instance, this means that *at the very least* each girl they occupy must be at or above their respective Bottom(). Furthermore, stand-offs need to be resolved in **b**'s favor always. Nevertheless, if these conditions are met, should the algorithm return true, then boy **b** has a strategy to get girl **g** or better (under the assumption players are allowed to make agreements, as per Definition 4.25 (agreements)): the matching determined when in line 4 **true** was returned is the final outcome for the boys in **b**'s coalition; **b** then makes the bitter sweet offer for this outcome to the other players – should all accept – and demands promises that they all punish the first non-cooperator in the sequence should one of them not accept.

One might run the above algorithm for all boys, binary searching the highest girl they could attain if all standoffs are resolved in their favor. Clearly if results differ for different boys, then a real standoff situation exists. Some of them need to back down. How to determine who would do this is again an interesting question: maybe the ones which would get hurt the must under a don't cooperate / don't cooperate situation. Complications clearly arise in answering this different question.

Note that Algorithm 4.26 only produces meaningful results if the boys are allowed to communicate before the beginning of play, making and responding to Ultimatums and proposals. Note that in this instance, the strategy implied by Algorithm 4.26 also works with mere static strategy profiles – a profile is chosen as either the cooperation matching, or the punishment matching for the first non-cooperating player. If, however players are not allowed to negotiate before the play, a dynamic strategy profile might or might not exists for each coalition player to punish the first non-cooperator: it depends on whether the punishment matching is reachable from the Game_State the coalition boys find themselves in after the first iteration of the Dynamic Game Model 4.6. Thus the situation when a boy is lying when accepting to be part of a coalition is not handled.

Unfortunately, Algorithm 4.26 (expanding wrath approach) runs in super-polynomial time.

**Theorem 4.27 (running time of Algorithm 4.26)**
The total running time of the Algorithm 4.26 is no more than O(**n!** * **n**$^5$ * log **n**).

*Proof*: The depth of the recursive calls of the routine is at most **n**, with each potential sequence of boys appearing at most once. Thus, there are at most **n!** potential stacks of recursive calls of routine IsOutcomeFeasible(), each being comprised of at most **n** players. The running time at each level in the stack is the total time taken by lines **2-6**. Line **2** can be made to take O(1). Line **3** can take no more than O(**n**$^3$) using the maximum matching algorithm based on Ford-Fulkerson [14]. Line **4** also takes no more than O(**n**$^3$) once the input graph has been constructed (which takes at most O(**n**$^2$) if done efficiently, or O(**n**$^3$) naively). Line 5 is executed at most **n** times. In line **6**, one can use binary search on the set size to determine the maximal **Sx**, running a maximal matching algorithm at each iteration, for a total running time of O(**n**$^3$ * log **n**). Line 6 is also executed at most **n** times, for a total of O(**n**$^4$ * log **n**). Thus, the total running time of lines **2-6** is O(**n**$^4$ * log **n**). There are at most **n** executions of these lines in a particular stack of recursive calls, for a total of O(**n**$^5$ * log **n**). The total number of stacks is no more than **n!** so the total running time is no more than O(**n!** * **n**$^5$ * log **n**).
|
*Discussion*: Note that this worst-case bound for the running time would make the algorithm feasible up until **n** = 12 for an ordinary computer. Not that much. It provides, nevertheless a useful framework to reason about threat making and a starting point for future refinements.

Note that the Algorithm 4.26 (expanding wrath approach) can produce false negatives: A boy might still be able to get a girl for which the algorithm returns false: for example some boy for which the set **Sx** is empty when he is considered in the algorithm, might actually not be able to get any of his higher preferences than some set of girls controlled by the coalition, due to plays by the other external players (e.g. his first preference prefers another external player who also prefers her first). Thus, he could be coopted by the coalition even though no such option is revealed in line 6 of the algorithm. Furthermore, the coalition outcome might never be externally stable, but no external boy might be interested to veto the coalition: if Example 4.9 (devilish alliance), is extended to include another boy, **b11** who is the first preference of all the girls and who prefers girl **g11** tops, then the algorithm will terminate (quite quickly) with a false negative for any other boy.

Running the Algorithm 4.26 (expanding wrath approach) for **b4 – g1** for Example 4.9 (devilish alliance), would yield the desired coalition, under the permutation <**b4, b9, b10, b2, b5, b7, b8, b6, b3, b1**> for example. Note that running it for **b1 – g1** returns (under proper choice of girls' preferences) **false**: the maximal alliance **b1** can muster is <**b1, b3, b6, b8, b7, b5, b2**> which is not externally stable and cannot promise anything of interest to **b4** to coopt him. While this is correct (**b1** cannot be sure he will get **g1**), it might still be that **b4** will settle for **g10** instead of exercising revenge over **b1** and ending up worse off, thus allowing **b1** to be coupled with **g1**. However, the algorithm properly reveals the situation that **b4** can offer **b1** something in exchange for cooperation, while **b1** cannot offer **b4** anything he cannot take for himself.

To try to resolve the issue of false negatives, the following need to be addressed:
- The girl **g1** which is currently determined in lines **6-7**, must be the one right above the first girl the candidate boy **b1** could get if he ignores the coalition (which becomes hostile to him). Determining such a girl is no easy task: he may or may not be helped by the other external players in getting a good outcome!
- In determining external stability of a matching, one should ignore players which have no interest in vetoing a coalition. This is also no easy task: An external player may be coerced or outright forced by other external players (an alternative coalition) to exercise such a veto! Even if this is not the case, determining that he can get a better outcome by not vetoing (or that he can be dearly punished for a veto) is no easy task!

We hope the concepts and results introduce in this section shed some light into the important issue of strategic play in a stable marriage game when threats and dynamic strategies are permitted. Since as far as we know this topic has not been addressed so far in literature under the assumption set and model of this section, we leave further developments and refinements for future research.

In particular, we leave several open questions which are explicitly stated in Section 8.

# 5. FURTHER POSSIBILITIES

In this section we discuss a few other forms of strategic play might affect the potential outcome for a player. Except threat-making, is there any other way in which he can hope to improve his outcome? We also briefly state how such situations might come into being.

There are six situations we analyze which can contribute to altering a boy's outcome:
 1. Removing a boy from play: The set **B** decreases by the removed.
 2. Adding a boy to play: The set **B** increases with the added boy.
 3. Removing a girl from play: The set **G** decreases with the removed girl.
 4. Adding a girl to play: The set **G** increases with the added girl.
 5. Convincing other players to side with them in coalitions, without making any offers to improve their matching or threats to degrade it.
 6. Controlling a set of drone players: players which are not interested in their own individual matching, but instead are controlled by some other player which dictates their play.

In this section we deal with existential questions only: can such situations generate a better or worse outcome for a particular player? We ask these question in the framework of four different models:
- Naïve Gale-Shapely
- Top-trading-cycles matching as per [10 - Cheating by Men in the Gale-Shapley Stable Matching Algorithm, Chien-Chung Huang, 2006]
- Coalition stable matching as per Section 3
- General situations with threats and dynamic strategies as per Section 4.

If the outcome can be improved by the specific situation we ask, what actual such operations would need to be performed (or can be performed) for some boy **b** to end up coupled with some higher option girl **g**. It is obvious that in the first four situations the operation can have no effect: for example, if the acted on boy is the last preference of all girls, respectively the acted on girl is the last preferences of all the boys.

Removing a boy from play

Under the naïve Gale-Shapely play, removal of a boy can only improve (or leave unchanged) the outcomes for the others. Removal of a boy implies removal of the proposals he makes. This in turn can only result in potential removal of other proposals by others as their parent set becomes empty, under Definition 3.27 (information about elements). To model the effects of removing a boy, one might just run Gale-Shapely again over the remaining boys. Trying to maintain the tree of proposals under deletion does not immediately lead to a better running time algorithm.

Clearly removing all boys **b1** which propose in Gale-Shapely to **g** and beat **b** there will allow **b** to get **g**. However, it may not be necessary to remove all such boys (or any of them at all!). Consider the situation in Example 3.29 (inferno of boy b). All players attain optimal outcome if boy only **b** is removed (while they are all beaten at this optimal outcome by n-1 players which, if boy b is not eliminated, all propose). Same applies if boy **b1** is removed in that example. Furthermore, if we were to extend Example 3.29 (inferno of boy b) with another pair say boy **b'** and girl **g'** such that **b** has **g'** as his first preference but is beaten there by **b'**, then removing **b'** (which does not even propose to **g**, **g1**,…,**gn**) allows **b1** to get girl **g1** for example.

Under the top-trading-cycles method, removal of a boy might improve the outcomes of a player by improving the Gale-Shapely he is considered to have control over and thus be able to trade, as discussed above. However, it might as well degrade his outcome by removing a potential trading partner. Consider Example 3.1. If boy **b2** and **b7** are removed, then boy **b4** no longer has a trading cycle to his first preference, **g3** and is stuck with **g2**.

Under the coalition stable method of Section 3, removal of a boy can definitely improve the outcome for the others, since it can improve the Gale-Shapely matching the play of each iteration (in particular the first), as discussed further above. However, it can also degrade it. In Example 3.1, if **b7** was to be removed, **b6** would become a fist-iteration hopeless man (alongside **b5**), his outcome degrading to **g6**.

Under the general game model when threats are permitted, clearly removing a threat yielding boy can improve the outcome for some – as it is easier for them to construct externally stable matchings: consider removing **b4** from Example 4.9 (devilish alliance). Just as well, removing a boy might also degrade the outcome for some of the others: he is no longer available as a potential valuable member of a coalition. In Example 4.9, removing **b9** would make **b4** unable to kick **b1** out of **g3**, his second option. He still has the option of kicking **b1** from the fought-for **g1**. This option also vanishes if **b2** is removed instead of **b9**. Note that in this latter case, the outcome for **b9** is also degraded (since **b4** will play to his second choice, **g10**, being unable to deter **b1** from occupying his beloved **g1**). Note that **b2** was considered in Example 4.9 a naïve player and was not explicitly required to be part of the coalition, merely to try to get as good an outcome for himself as he could.

How might a boy be removed from play in real-life? Most economic models allow for several situations, like:
- The player not being knowledgeable about the existence of the girls under dispute (i.e. playing the game knowing just a subset of the available girls), thus having the effect of being as-if absent.
- The player being offered incentives (or threatened with disincentives) outside the game model to refrain from participating in a particular instance of the game. Regulatory requirements might apply making it risky for a player to engage in the game (for example a US small-sized actor to have business deals with a

- Caspian See, Middle-eastern country). Incentives offered within the scope of the game, can be modeled as offers to join a coalition.
- Offering him another girl, outside those under dispute, to leave the game (potentially a girl which is not known to the other boys and, as such, does not get better proposals).
- The prospect of future repeat plays of the game including the same set of players which asked him to leave that particular instance, which may, in turn, do the same for him in the future (or have already done so in the past).

Adding a boy to play

Adding a boy has the opposite effect on the game instance as removing him from the new game instance has. Except for the naïve Gale-Shapely, we discussed above that removal of a player can either benefit or hurt other players. As such, adding one can either hurt or benefit others. For the naïve Gale-Shapely case it is also trivial that adding a boy can degrade the outcome for others.

How might a boy be added to play in real-life? Most economic models allow for several situations, like:
- A player priory not being knowledgeable about the existence of the girls under dispute (i.e. playing the game knowing just a subset of the available girls), is made aware of them.
- The player being offered incentives (or threatened with disincentives) outside the game model *to participate* in a particular instance of the game. For example, he could be incentivized to participate by including the girl he is already coupled with in the game. If he does not participate, he risks losing that matching.
- The player is being made eligible to participate in the game (for example, by paying his entrance fee, or facilitating him getting a required letter of credit).
- The player is raised up the preference list of some girls, by altering the underlying criteria pertaining to him, based on which the girls decided the preference ranking (for example, by increasing his wealth or having him become part of a more affluent entourage).
- The prospect of future repeat plays of the game including the same set of players which asked him to *join* that particular instance, which may, in turn, do the same for him in the future (or have already done so in the past).

Removing a girl from play

Removing a girl might leave fewer girls than boys. For this reason, by convention, removal implies substituting the removed girl with another fictitious girl, which lies at the end of the preference lists of all the boys. If any boy ends up coupled with the fictitious girl, he is in effect uncoupled. There can be more than one fictitious girl. They all represent the situation where the respective boys are uncoupled. The preferences of the fictitious girls themselves are irrelevant: should one be under dispute; the loosing boy always has another (equally fictitious) girl which is uncoupled. The assumption is, of course, that boys will seek to avoid remaining uncoupled, by proposing to a fictitious girl.

Under the naïve Gale-Shapely play, removal of a (non-fictitious) girl will definitely degrade the outcome for at least one boy – the one who would end up with her –, if the number of players is less than that of (non-fictitious) girls after the removal. Removal of a girl can also degrade the outcome for others: e.g. removing the first preference of a strong player (which would remain coupled with her), might trigger a long play, degrading the outcome of several boys. For illustration, consider that in Example 3.29 (inferno of boy b), that boy **b**'s first preference (where he would play unchallenged) was removed and replaced by the fictitious girl **g**. The result is catastrophic for all players. Under naïve Gale-Shapely, removal of a girl cannot improve the outcomes for any boy: removal only has the effect of more proposals being made to the remaining girls, which can only increase (or leave unchanged) their temperature.

Under the top-trading-cycles method, removal of a girl can clearly degrade the final outcome as it can degrade the Gale-Shapely outcome which is used to determine which girls are controlled by which boys, which can then be traded. However, it might also improve the outcome for some: e.g. a boy whose first preference is removed, causing him to remain coupled in Gale-Shapely with his third preference, will be very interested to trade her for his second preference, thus potentially becoming part of a trading cycle (which he was not when he got his first preference directly).

Under the coalition-stable matching, removal of a girl can clearly degrade the outcome for some boys, as it negatively impacts the Gale-Shapely outcomes (e.g. the outcome of the first round), as discussed further above. Removal can also positively affect the outcome for some boys: a boy whose first preference is removed, can oust another player from his third preference, causing him to be that round's hopeless man; once that man is no longer considered (being hopeless in that round), the first player can remain (in the next round) with his second preference, which was occupied by someone having the latter's third preference as first preference and who, before the freshly hopeless man was eliminated, would not have been able to remain with her (due to the freshly hopeless man, which was not hopeless before).

Under the general game model, removal of a girl can both degrade the outcome for a player (e.g. the first preference of a boy first preferred by all girls is removed) and improve it: a boy external to some coalition may more easily be subjected to a bitter-sweet threat, consisting of an offer to remain coupled with a girl over which belongs to a set of girls controlled by the coalition, since he has fewer outside options.

Note that removing a girl can sometimes be viewed as adding the boy who priory remained coupled with the removed girl.

How might a girl be removed from play in real-life? Most economic models allow for several situations, like:
- Keeping the players from being knowledgeable about the existence of the removed girl, thus having the effect of her being as-if absent.
- The girl being offered incentives (or threatened with disincentives) outside the game model to refrain from participating in a particular instance of the game. The same can apply to boys being affected in this manner to abstain from playing at that girl.
- Jamming the "communication channels" to the girl, effectively preventing her from receiving or responding to proposals.
- Degrading the characteristics of the girl so that she becomes least preferred by all the boys.
- Offering the girl another boy, outside those taking part in the game, to leave the game (potentially a boy may not be aware of all the other girls in play; or may be aware of just a subset of the boys in play – which would make him wrongly estimate his outcome under a coalition/threat strategy). Note that a girl might not be interested to leave the game since, should she not leave, she might get an even better outcome than the promised boy (who can still propose to her). However, if that boy could get himself a better outcome than her should both part-take in the game, then she might as well accept him and leave the game.
- The prospect of future repeat plays of the game including the same set of players/girls which asked her to leave that particular instance, which may, in turn, do the same for him in the future (or have already done so in the past). Note that when the removal of a girl has negative impact on the outcomes for some players, it must have positive impact on the outcome for some girls.

Adding a girl to play

Adding a girl has the opposite effect on the game instance as removing her from the new game instance has. Except for the naïve Gale-Shapely, we discussed above that removal of a girl can either benefit or hurt other players. As such, adding one can either hurt or benefit others.

For the naïve Gale-Shapely case it is also trivial that adding a girl cannot degrade the outcome since she is either ignored, or she is the object of some proposals which might in the end leave other girls at a lower temperature.

How might a girl be added in real-life play? Most economic models allow for several situations, like:
- Making the players in the game aware of her existence, assuming they were not aware of it before.
- Allowing some boy she was priory coupled with to renounce her and become uncoupled (perhaps to join the game), thus also leaving her uncoupled and incentivizing her to join the game. In real life economic situations this can happen for example at the expiration of the term of a contract (e.g. for leasing

- real-estate, for doing consultancy work, etc.).
- Improving the characteristics of a priory "as-if fictitious" girl so that she becomes considered by the boys (e.g. renovating a house, having a renowned professor join a school, etc.).

Coopting players in coalitions without incentivizing them within the game

We refer to situations when a player is no better - no worse off by joining a coalition. This can happen with the accomplices of the top-trading-cycles method described in [10 - Cheating by Men in the Gale-Shapley Stable Matching Algorithm, Chien-Chung Huang, 2006], with the hopeless men of each round in the coalition stable matching Algorithm 3.22 – dropping Assumption 3.10 (self-interest and no taking of sides), or, in the case of the general game model, with a player who cannot be coerced or incentivized by a coalition (already getting a better outcome than the coalition might offer), but who voluntarily joins it, engaging in threat-making over other players. Clearly there can be no relevant situations in naïve Gale-Shapely since that game model allows no coalitions or decisions.

Under the top trading cycles method, clearly all accomplices need to support the coalition of the boys who are part of a trading cycle, thus effectively joining it. Since they are not part of any cycle themselves, they do not get a better outcome. However, they don't get a worse outcome either by supporting. It could be nevertheless, that they might have a preference for some other boy to get as good an outcome as possible. In this case, certain trading cycles (over which the accomplices have legitimate direct veto power) cannot materialize.

Under the coalition stable matching, in each round of Algorithm 3.22, the hopeless men might also have a preference to help some other remaining boy, in exchange for them playing directly to their final match in that round.

Note that in both situations, such side-taking is equivalent to threat-making, except that threats are enabled only by legitimate direct vetoes, not any kind of direct vetoes. The complications revealed by Section 4 can easily spur: stand-offs might emerge between different players which old direct legitimate veto over each other's preferred matching for their friend player; in Algorithm 3.22, players eliminated earlier can threaten later rounds players not to favor any other boy then their friend, under threat of not playing directly to their coalition stable girl – they might also collude in doing this; in the top-trading-cycles method players might be threatened that their own trading-cycles would be vetoed if they do not support a better outcome for some other player. In effect, reducing the spectrum of direct vetoes to include just legitimate direct vetoes does not seem to simplify the general case much. Enabling threats only to legitimate vetoes can, in Algorithm 4.26 (expanding wrath approach) for instance, be modeled by disallowing edges to girls below a player's matching (under Algorithm 3.22 or Gale-Shapely).

Under general play, a further situation might arise: A player might join a coalition without being coerced or offered any incentive (e.g. a player who gets his first choice, might still join a coalition). In this situation, more sets of girls can fall under the control of the coalition, potentially allowing it to grow further, for example under Algorithm 4.26 (expanding wrath approach). Note that in the general model however, a player joining a coalition without consideration might not be considered credible by the others. Under Algorithm 4.26 for example, he might be required to make threats of plays to girls below what he would get if he didn't join the coalition. Since he has no incentive to join the coalition in the first place, such threats could be considered non-credible. The issue can be resolved however if he is allowed to make threats only of exercising direct legitimate vetoes – only playing to girls at or above the girl he would remain coupled with if he didn't join. Note that these independent players might not join any coalition at all. They may simply play as if they were naïve players and still end-up with one of their top preferred girls – they cannot be deterred from all by any rational coalition.

How can players be incentivized to join a coalition which does not offer them any improvement (but does not degrade their outcome either)? Among others:
- Players could have a preference for someone introducing them to the game (e.g. by revealing certain priory unknown girls or other players).
- Players could be offered side-payments. Note that in this case, in the models we examined we required that a player is interested foremost by his resulting matching. Side-payments could thus only be used to break ties among different strategies all of which produce the same outcome in the game for the player. One relevant model with general transferable utility has been analyzed in [13 - Designing a strategic bipartite matching market, Rahul Jain, 2007].
- Players might consider the prospect of future repeat plays of the game including the same player(s) who asked for support in that particular instance, who may, in turn, return the favor in the future (or have already done so in the past).
- Players could be friends, thus favoring each other if that does not hurt their own outcome.

Drone players

A slight variation to the game model is to allow the existence of players whose play is entirely dictated by another player (under a static or, respectively, dynamic strategy profile). Introduction of drone players may be accompanied by the introduction of extra (potentially fictitious) girls so that the total number of players does not exceed the number of girls.

Under naïve Gale-Shapely, introduction of extra players – be they drones or not – can only degrade the outcome, as discussed priory in this section.

Under the top-trading-cycles method, a boy is interested in one of his drones ending up coupled with his target girl, so that they could later switch. If that is not possible, he could be interested that his drones complete a trading cycle which includes him and his target girl: namely that one of them occupies a girl better preferred by another boy, who occupies a girl better preferred by yet another and so on, culminating with the target girl. Note that is not useful to have more than one drone on a trading cycle since, considering the trading cycle starting from the target girl, once it comprises the first drone, that drone could as well occupy the girl held by the controller, thus leaving the target girl for him, completing the cycle. The rest of the drones might still be useful in pushing other players on that trading cycle from their higher-up preferences, thus making the cycle possible (which may, in the end, not necessarily include a drone).
Note however, that under this approach, the drone which is part of the cycle (if any), might be better preferred by the girl the controller ends up with. In this situation, an automated system might not even permit that the drone does not play first to the girl the controller would trade to the drone, thus making the trade no longer possible. This might be circumvented by the introduction of a special girl (least preferred by all others), the controller can end up with and trade with the drone. However, this might prevent the controller from ousting some of boys which are part of the trading cycle, from their better preferences.

Under the coalition stable matching, introducing drones can be useful. At the very least they can be regarded as players who never play to the controller's target girl. This can sometimes be sufficient in itself as discussed priory in this section, regarding Adding Players. Sometimes however drones should play to the target girl of the controller – if he was kicked-out of there, they may need to also kick out the kicker so that he becomes a hopeless man in an earlier round than the controller. In fact, in the round in which the controller would become hopeless under Algorithm 3.22 without existence of drones, the controller should play (successfully!) to a girl occupied by a drone who must trigger a play which kicks-out the player who kicked him out from the target girl (ideally making him hopeless). This way, he does not risk being made hopeless by one of his own drones who beats him at the target girl. In fact, if a boy has at his disposal a drone which would beat any contender at his target girl, he could always "wake up" that when kicked out from this girl. This assumes, of course, that there exists a girl which ends up occupied by the drone where he beats the drone. Such a girl could be introduced artificially, by Adding a girl: some girl least preferred by all others.

Under the general play, introducing drones can be provide a great advantage to the coalition of its controller by potentially causing new sets of girls to fall under the control of the coalition or which can be denied to some other boy in case of non-cooperation, in the course of Algorithm 4.26 (expanding wrath approach) for example. The drones have the added advantage that they can settle for any outcome, thus not being an impediment to any matching which includes them. Algorithm 4.26 could be extended to include drones by not considering them part of the players when determining the existence of a matching in lines 3 and 4, but considering them otherwise.

Note that in all cases where drones are used, it could be useful if they are allowed to just "hover" unmatched until their controller "wakes them up" to make some proposal. This however is not currently permitted by either the Dynamic Game Model 4.6 or the Simplified Game Model 2.10.

We conclude this section by leaving open the questions of finding an efficient algorithm to determine, for set of optional players (and potentially another of optional girls), which should be added and which excluded by a player in order to gain an optimal outcome. The brute force method would imply solving the instance for each potential choice, adding an exponential factor to the running time.

# 6. VARIATIONS TO THE GAME MODEL

In this section we discuss how several slight variations to the game model can be taken into account.

<u>Girls with more than one slot</u>

There are cases (for example in the student-program matching) when a girl can remain coupled with several boys.

One way this could be modeled in the problem instance would be to group all slots for the same girl adjacently in the preference list of the boy corresponding to the position of that particular girl on his list. For example, **Pb1**= <**g1**-1,**g1**-2,**g1**-3,**g2**,**g3**-1,**g3**-2,**g4**>. If in the final matching a boy is coupled with any slot of a girl, then he is said to be coupled with that girl. The girl will have the same order of preference over the boys at each slot. When several boys compete for the same girl, they actually compete for a particular slot. The loser can the try to get the next slot and so on. By convention we could require that slots are grouped in the same order in all the boys' preference lists. However, this is not required. In this situation a boy is considered indifferent among the slots of a particular girl.

One can quickly note that running Gale-Shapely over this adjusted preference list produces a stable matching (no only over girl slots but over the girls themselves): the temperature of all slots of a girl will be above a certain boy if he proposes to her and ultimately does not get any of her slots. Furthermore, this stable matching is still man-optimal (among other stable matchings), by the same arguments used for Gale-Shapely.

The top-trading-cycles method would also work unaffected over the new formulation of the problem: a trading cycles of girls by boys can be formulated as a trading cycle of the underling girl slot by boys.

The coalition stable matching method also requires no further adjustment. All results still hold.

Under general play however, one has to consider the adjustment as with any case with indifference by boys, as is discussed further in this section.

Note that this resolution to the indifference problem has the potential drawback of increasing the size of the problem: each girl may have up to **n-1** slots (having more than that would make her able to accept all boys). Thus, the number of girl slots under the adjusted problem instance can grow to O(**$n^2$**), which would adversely affect the running times of various algorithms. To circumvent the problem, one can instantiate slots only when they are effectively being used (about to be occupied by a boy who was kicked-out or refused at all prior slots), up until the slot count limit of the particular girl. In this manner there will not be more slots than man, with the naïve Gale-Shapely, top-trading-cycles method and coalition-stable matching Algorithm 3.22 maintaining their O(**$n^2$**) running times. However, in the general play when threats are permitted, all slots count: a boy may seek refuge in a then-unused slot of a girl and a threatening player has to threaten him from all slots of a girl. Note that in all cases, just the top **n** preferences of slots (be they for the same or different girls) for each boy matter: he will definitely get one of those, no matter what. Thus any algorithm which is bound to walking the preference lists of boys will have an input of size O(**$n^2$**).

Indifference in preference lists of boys

One can extend the idea of girl with multiple slots to generally allowing indifference in the preference lists of the boys. Note that unlike in [2 – Stable Marriage and indifference, Robert W. Irving, 1994] here we require that the preference lists of just the boys allow indifference, not that of girls too.

Choosing an arbitrary order for the cases which give rise to indifference can be a useful way to generate an adjusted problem instance.

As with the case above, in this situation Gale-Shapely will produce a stable matching. However, considering that some boys may be indifferent among several girls they could end up coupled with, choosing a particular girl from this set (effectively permuting the order in which the indifferent choices appear in the positions they occupy within the boy's preference list) may make her unavailable to another boy who does not have the luxury of having several equally preferred choices to her that he could get. But choosing a different girl among those equally preferred, might make another boy worse off!

Using the top-trading-cycles method to improve a matching can, of course, yield a better result for some boys, however with all the drawbacks applying this method in the usual case (with no indifference) has: the resulting matching may not have the support of players with legitimate direct veto power which can attain a better matching for themselves by including some other set of players in their coalition (often times just a subset).

Using the coalition stable matching Algorithm 3.22 still yields some valid result; however, the order in which a boy proposes to the girls at a certain level of preference affects the end result for the other players.

**Example 6.1 (order matters with indifference)**

Say the preference lists of the boys are as follows (with parenthesis being used to signify indifference).

Preference of boys, in descending order (only relevant ones):

| Pb1 | g1, g3, g2 |
|---|---|
| Pb2 | g1, g2, g3 |
| Pb3 | (g2, g3), g1 |

Consider the following two alternative plays:

Round b1: b1→g1.
Round b2: b2→g1 | b1→g3.
Round b3: b3→g3(b1) | b3→g1 | b2→g2.

Here just **b2** is a hopeless man (**b1** is not because he defeated **b3** at **g3**), ending up at **g2**. After being eliminated, in the next round **b1** will get **g1** and **b3** will get **g3**.

And:

Round b1: b1→g1.
Round b2: b2→g1 | b1→g3.
Round b3: b3→g2.

Here both **b1** and **b3** are hopeless men, ending up at **g3** and **g2** respectively. After being eliminated, in the next round **b2** will get **g1**.

|
*Discussion*: Notice how the difference in the order in which **b3** exercised his proposals at a certain level of preference affected the outcome for the others. Also note that this order can only matter when a boy ends up at that level of preference: if he drops below it, he will surely have made all proposals at the prior level in that round's Gale-Shapely.

Thus, in the coalition stable matching, Assumption 3.10 (self-interest and no taking of sides) might no longer help in some situations: Proposing one way would disfavor one player, while proposing the other way will disfavor the other player.

We might be able to improve the matching obtained by Algorithm 3.22 for some player,

by trying to find an alternating path consisting of that player being matched to some higher preference and all other players matched only to preferences at the same level of indifference as before. This can be determined with any walk of the bipartite graph, thus in time at most O($n^2$) for any single boy. Note however that improving the matching for two boys simultaneously might not be possible – thus some side would again have to ensue so as to determine for which one the matching will be improved.

Under general play, there are two complications added by indifference:
- When a boy is threatened from a girl, he must in fact be threatened from all girls at that level of preference;
- When a boy is offered a girl, the offered girl must not be from the same level of preference as the boy would achieve without the assistance of the coalition.

Algorithm 4.26 (expanding wrath approach) can be easily adjusted to handle both these complications, by restricting offers to the first girl in set Sx above the level of preference of the first girl outside it. However, these adjustments may cause the algorithm to produce false negatives more often (as a boy may not be able to get any girl at a particular preference level without the help of the coalition, even though the coalition alone cannot oust him from all such girls).

Boys not knowing about all the girls in play

Situation may often arise in real life situations when some boys are not aware of the entire set of available girls (e.g. a contractor may not be aware of or have access to all potential customers).

In this circumstance, the affected boy effectively never proposes to a girl he is not aware of.

We can impose this condition by adjusting the problem instance as follows: up to **n** fictitious girls are added: these girls are least preferred by all boys among all other girls (it does not matter what the relative order of preference among them is – a boy ending up coupled with a fictitious girl is considered uncoupled). Then, the preference lists of boys are adjusted (based on the true preference list) as follows:
- All girls whom the boy is not aware of are removed from the preference list.
- The preference list is then compacted by eliminating empty positions;
- All free position up to the **n**-th in the boy's preference list are occupied with fictitious girls.
- After the fictitious girls, all girls removed initially from the list (the girls the boy is not aware of) are re-added (maybe in their actual order of preference).

Also up to **n** fictitious boys will need to be added, all least preferred by all the girls (including fictitious ones).

Since there are only **n** non-fictitious boys, a boy will surely remain matched with one of his top **n** preferences. Thus, the reminder of his list is irrelevant.

Note that this can at most quadruple the size of the problem instance.

Naïve Gale-Shapely clearly will work in this case. Top-trading-cycles method would have to exclude the fictitious boys (or these boys could be made to prefer the fictitious girls tops); otherwise it works without modification. Same applies to the coalition stable matching Algorithm 3.22.

Under the general play, the running time of algorithms may be adversely affected (since Algorithm 4.26 for example runs in time exponential in the size of the problem instance). However, the fictitious boys can never be useful to be included in coalitions – they have no threat yielding power. Nevertheless, coalitions which offer them "better than fictitious girls" might temporarily be considered – point out the weakness in terms of running time of Algorithm 4.26.

<u>Boys not knowing about all other boys</u>

Boys may not be aware of all the other players which exist in the game. This situation might be compounded with the one just discussed, where they are unaware of some of the girls part-taking in the game.

Can this affect game-play?

Under the naïve Gale-Shapely play, clearly no. A player just might be surprised in the end to learn that some boy he didn't know was in the game occupied one of his preferences, or (directly or indirectly) the preference of some other boy who occupied his preference.

Under the top-trading-cycles method – besides the possibility of the Gale-Shapely starting point being different to what a player expected – players which are not aware of each other (and are not introduced to each other) cannot trade with each other directly. A boy can be generally considered aware of what boy occupies a girl he knows was in play. As such, players unaware of each other may are typically also unaware of the girl occupied by the other. But even in this case, other players on a potential trading cycle might "intermediate" the trade – offering a "new and better" girl to each player who does not know the other party. However, trading cycles in which there is not a single boy who knows about all the others cannot materialize. Verifying the existence of a trading cycle could be done in $O(n^2)$, by doing a DF walk for each boy, only on other boys known to him (for whom he can serve as an intermediary in trades).

Since the coalition stable matching essentially consists of running Gale-Shapely on smaller and smaller problem instances, with no intervention by the players, the only potential issue is that, again, they may be surprised (negatively or positively) at the end result being different than what they anticipated.

Under general play, boys a coalition is not aware of cannot be coopted in the coalition. Thus the coalition is potentially deprived of a useful advantage. Much more troublesome, strong unknown boys might emerge to veto (perhaps legitimately) a matching the coalition believed was externally stable.

<u>Girls as players</u>

In this paper, we made Assumption 2.1 (girls are robotic players). We argue that for the situation of Section 3, even if the girls are allowed to *also* propose this will not affect the outcome, since the boys, forming a grand coalition, can reject proposals by the girls outside their man-optimal matching. However, this argument assumed that the girls are still obligated to accept a proposal they receive (or get accepted) rather than remain uncoupled. Boys on the other hand can reject proposals they receive outside those themselves make.

There are several potential variations to the game model which present themselves when girls are allowed to have decision points, thus becoming players:
1. *Girls being allowed to make proposals*. In this variation, players from each group can have a single binding outstanding proposal and several received ones. They must then start rejecting proposals (which can never again be made), leaving out just the outstanding proposal and at most one received proposal. If their outstanding proposal gets rejected, they can make another one, or choose from one of the received proposals not already rejected. Since incoming proposals are guaranteed, one might think that it is optimal for a player to reject all received proposals, save the best. However, situations could occur where the outstanding proposals form a cycle in the bipartite graph of the game. Clearly the proposals on a cycle can never be all accepted. But which one (or which ones) get to be considered rejected? There is no fair answer in this case.
We could also allow players to have more than one outstanding proposal. In this case, situations might arise where several outstanding proposals are accepted – in this case, the player would be unable to honor all of them. We could, of course, allow that a player changes his mind about an outstanding proposal, canceling it – as sometimes is the case in real-life play. The classical model of the stable marriage problem can be viewed as girls having such outstanding, cancelable proposals to all boys, with them canceling some lesser preferred one as they themselves get proposals by the boys. In real life situations it is sometimes the case that a player will make or accept a proposal tentatively – just waiting to get a better one. Furthermore, a player might lie to a proposer that his (her) offer was accepted so that he (she) is kept available as "worst case", up until some timeout. This can cause the lied to player to ultimately get a worse outcome than if he hadn't believed the acceptance of the offer, since he could have refused some offers in the interim, which are no longer available when he ultimately becomes free as the counterparties might have already entered binding arrangements with someone else. These kinds of situations sometimes arise in real-life play.
2. *Girls being allowed to reject proposals better than their then-current partner.* In this case we may require that once a proposal has been rejected, such a proposal cannot be made again at a later time – a girl cannot change her mind about preferring a boy from whom she rejected a proposal. In some real life cases (for example in contractor – client matching) this can clearly not be the case. However, often times it is. Furthermore, if we allowed rejected proposals to be later accepted, some boys might keep proposing to the same girl (e.g. their top choice) over and over in the hope the girl changed her mind. A standoff would occur as to which boy is the first to propose to another girl. A special case of this variation is when:
    o *Girls are allowed to just permute the preference order for accepting proposals.* In this special case girls have to state up-front an order of preference under which they accept proposals (which does not have to be the actually order of preference for the final matching). This is can be viewed just as girls "falsifying" their preference lists in all algorithms we presented in this paper. We call this situation as *girls having static strategy profiles*.

If girls are allowed to reject proposals even when uncoupled, then the game can turn into one of ultimatum between boys and girls: with boys proposing just their man-optimal coalition-stable outcome and girls accepting only some coalition-stable matching of their own (perhaps the man-optimal coalition stable matching for the converse problem instance – when the roles are switched between the two sets). If boys are forced to keep proposing if refused, then girls can force any outcome they collectively (all of them) choose.

The situation where girls are allowed to rejected initial proposals in static strategy profiles can easily be modeled by adding one fictitious boy **b'** for each girl **g**. The special boy **b'** will have girl **g** as his only preference. Thus, girl **g** will be able to choose **b'** instead of becoming coupled with any of the real boys. If we require that the number of boys and girls remains the same, we can also add another girl **g'** (least preferred by all non-fictitious boys) which will **b'**'s second choice and for whom **b'** is her first choice. In this circumstance it could occur that some boys and some girls remain uncoupled (i.e. the final result involves them being coupled with fictitious partners).

In analyzing strategic plays involving girls, the first question we ask is "can girls benefit from having a static strategy profile other than the naïve one?". Clearly if they collude among themselves in situations parallel to the one constructed in the paragraph above, they could get better outcomes. Even without being allowed to reject a first proposal, a coalition of girls might decide to refuse the other's agreed on partner to the detriment of some boy outside their coalition matching (if, of course, there exists such a boy who proposes before or simultaneously with the first boy in the coalition matching). If for all boys in the coalition matching, the boy's preferences between the first girl who takes part in the coalition and the girl in the coalition to whom he is "promised" consist only of coalition girls (or other girls who reject him), under naïve Gale-Shapely play by the boys, the coalition outcome of the girls will be attained.

It might be interesting to have an algorithm for determining potential coalitions of girls, along with associated outcomes for when boys are assumed to be robotic naïve Gale-Shapely players. We leave this open for future research, maybe using [4 - Ms. Machiavelli and the Stable Matching Problem, David Gale and Marilda Sotomayor, 1985] as a starting point.

One very, very interesting case, which is sometimes easily overlooked is that of coalitions consisting of both boys and girls. In this paper we focused on generating a good outcome for some boy or boys, so we will keep this focus and ask: "might it be useful for a coalition of one or more boys to attract some girls to it?". However, the question might be posed conversely too: "might joining some coalition of girls be useful for a boy?".

We discuss just the case where girls have static strategy profiles. The case when girls can reject proposals based on the full Game_State is more general.

In engulfing a girl in a coalition, one first has to ask about what her motivation might be to join. In the scope of the game, we required that each player has the motivation to get as good an outcome as possible (with ties between strategies producing the same outcomes being broken in various manners). We can impose the same on girl-players.

Note that even one girl alone might benefit from rejecting a boy, even under a static strategy profile.

**Example 6.2 (lone girl benefiting by rejecting proposal)**

Say the preference lists of the boys are as follows.

Preference of boys, in descending order (only relevant ones):

| Pb1 | g1, g2, g3 |
|-----|------------|
| Pb2 | g2, g3     |
| Pb3 | g2, g1, g3 |

Preference of girls, in descending order (only relevant ones):

| Pg1 | b3, b1     |
|-----|------------|
| Pg2 | b1, b3, b2 |
| Pg3 | -          |

Consider the following two alternative plays:

Round b1: b1→g1.
Round b2: b2→g2.
Round b3: b3→g2|b2→g3.

Here just **g2** played naively, ending up with her second preference, namely **b3**.

Alternatively:

Round b1: b1→g1.
Round b2: b2→g1.
Round b3: b3→g2(b2)|b3→g1|b1→g2|b2→g3.

Here **g2** played strategically, refusing **b3** even though he was preferred over **b2**, ending up however with her first preference, namely **b1**.
|
*Discussion*: Notice that in the second play, no boy is better off, with two boys being worse off.

Can some boy or boys end up better off by including girls in their coalition, while also offering them better or no worse outcomes (that they could not get without cooperation of the coalition)?

The naïve Gale-Shapely play by boys allows for no coalitions including boys, since boys do not face decision points.

Under the top-trading-cycles method however, such arrangement could be attained.

**Example 6.3 (coalition of b2 and g2)**

Say the preference lists are as follows.

Preference of boys, in descending order (only relevant ones):

| Pb1 | g4, g5 |
|---|---|
| Pb2 | g1, g3, g4 |
| Pb3 | g1, g2 |
| Pb4 | g2, g3 |
| Pb5 | g5, g2, g4, g1 |

Preference of girls, in descending order (only relevant ones):

| Pg1 | b5, b3, b2, |
|---|---|
| Pg2 | b3, b5, b4 |
| Pg3 | - |
| Pg4 | b2, b5, b1 |
| Pg5 | b1, b5 |

Consider the following four alternative plays.

The first one is the naïve Gale-Shapely for all players.

Round b1: b1→g4.
Round b2: b2→g1.
Round b3: b3→g1|b2→g3.
Round b4: b4→g2.
Round b5: b5→g5.

Here all boys get their first preference, except **b2** who gets his second at **g3**. Girl **g2** gets her second preference **b5**.

The second play involves **b2** colluding with **g2** as follows (deviations from Gale-Shapely are in **bold**):

Round b1: b1→g4.
Round b2: b2→g1.
Round b3: b3→g1|**b2→g4**|b1→g5.
Round b4: b4→g2.
Round b5:
b5→g5(b1)|b5→g2**(b4)**|b5→g4(b2)|b5→g1|b3→g2| b4→g3.

Here **g2** played strategically, refusing **b5** even though he was preferred over **b4**, ending up however with her first preference, namely **b1**. Under this play, boy **b5** can trade **g1** (his 4$^{th}$ preference) for **g4** (his 3$^{rd}$ preference), under the trading cycle **g1|b5→g4|b2→g1**, which causes **b2** to upgrade from **g4** (his 3$^{rd}$ preference) to **g1** (his 1$^{st}$ preference). Any other trading cycle for **b5** would have to involve either **b1** (who gets his 2$^{nd}$ preference and cannot get his first one being occupied by **b2**) or **b4** (who gets his 2$^{nd}$ preference and whose first preference would be what **b5** also wants). Thus the trading cycle which gives **b2** his 1$^{st}$ preference, **g1,** would materialize under the top-trading-cycles method!

Note that both **b2** and **g2** have to collude to obtain this outcome. Just one of them deviating was not enough.

Consider the play where just **b2** deviated:

Round b1: b1→g4.
Round b2: **b2→g4**|b1→g5.
Round b3: b3→g1.
Round b4: b4→g2.
Round b5: b5→g5(b1)|b5→g2| b4→g3.

Here **g2** end up with **b5** (her 2$^{nd}$ choice, not the first as before) and **b2** cannot trade with **b3** for their common first choice, **b1**.

If **b2** does not deviate, **g2** has no opportunity to deviate since she receives a single proposal, namely from **b4**.

*Discussion*: In this example, once **b2** decided to deviate, it was optimal for **g2** to also deviate. However, the preferences could easily have been adjusted to make **b5** girl **g2**'s true first preference and **b3** her true second preference. In this case, cooperation was also required of her. She must promise that she will refuse her true first choice in order to get her second one, still improving over her at least third one if **b2** did not deviate in the first place. Philosophically note that **b2** can, under this coalition, get his beloved **g1** to whom he may be the last preference! In fact, **b2** may be the last preference of all girls, save for **g4** where he can be third last.

Note that the above collusion can be materialized merely by **b2** and **g2** falsifying their preference lists – this way the top-trading-cycles method as applied in [7] will produce the coalition result for **b2** and **g2**.

Can such a situation occur for the coalition stable matching under Algorithm 3.22? We have already seen in Example 3.24 (lies of b2) that a single boy can play strategically without the help of any girl. However, could the help of a girl he can reward prove necessary? The answer is yes.

**Example 6.4 (another coalition of b2 and g2)**

Say the preference lists are as follows.

Preference of boys, in descending order (only relevant ones):

| Pb0 | g4, g0 |
|---|---|
| Pb1 | g4, g5 |
| Pb2 | g1, **g3**, **g4** |
| Pb3 | g1, g2, g3 |
| Pb4 | g2, g0, g1 |
| Pb5 | g5, g2, g0 |

Preference of girls, in descending order (only relevant ones):

| Pg0 | b5, b2, b4 |
|---|---|
| Pg1 | b0, b4, b3, b2 |
| Pg2 | **b5**, **b3**, b4 |
| Pg3 | - |
| Pg4 | b0, b1 |
| Pg5 | b1, b5 |

Assume the players **b2** and **g2** submitted falsified preference list, switching the order of the bolded preferences.

Over the falsified preference lists, the outcome is as follows.

Iteration 1:
Round b0: b0→g4.
Round b1: b1→g4(b0) | b1→g5.
Round b2: b2→g1.
Round b3: b3→g1 | **b2→g4**(b0) | b2→g0.
Round b4: b4→g2.
Round b5: b5→g5(b1) | b5→g2 | b4→g0(b2) | b4→g1 | b3→g2(b5) | b3→g3.

The hopeless man is **b3**. He remains coupled with **g3**.

Iteration 2:
Round b0: b0→g4.
Round b1: b1→g4(b0) | b1→g5.
Round b2: b2→g1.

Round b4: b4→g2.
Round b5: b5→g5(b1) | b5→g2 | b4→g0.

The hopeless men are **b2** and **b4**. They remain coupled with **g1** and **g0** respectively.

Iteration 3:
Round b0: b0→g4.
Round b1: b1→g4(b0) | b1→g5.

Round b5: b5→g5(b1) | b5→g2.

The hopeless man is **b5**. He remains coupled with **g2**.

Iteration 4:
Round b0: b0→g4.
Round b1: b1→g4(b0) | b1→g5.

The hopeless man is **b1**. He remains coupled with **g5**.

Iteration 5:
Round b0: b0→g4.

The hopeless man is **b0**. He remains coupled with **g4**.

Iteration 6:
<there are no more boys remaining>
Thus the final Result = {**b0**-**g4**; **b1**-**g5**; **b2**-**g1**; **b3**-**g3**; **b4**-**g0**; **b5**-**g2**}, where the strategic player **b2** gets his first preference and also girl **g2** gets her second preference.

Now consider the iterations of Algorithm 3.22 if only **b2** deviated, but girl **g2** did not also alter her preferences.

Iteration 1:
Round b0: b0→g4.
Round b1: b1→g4(b0) | b1→g5.
Round b2: b2→g1.
Round b3: b3→g1 | **b2→g4**(b0) | b2→g0.
Round b4: b4→g2.
Round b5:
b5→g5(b1) | b5→g0(b2) | b5→g2 | b4→g0(b2) | b4→g1 | b3→g2 | b5→g0 | b2→g3.

The sole hopeless man is **b2**. He remains coupled with **g3**.

Iteration 2:
Round b0: b0→g4.
Round b1: b1→g4(b0) | b1→g5.

Round b3: b3→g1.
Round b4: b4→g2.
Round b5: b5→g5(b1) | b5→g0.

The hopeless men are **b3**, **b4** and **b5**. They remain coupled with **g1**, **g2** and **g0**

respectively.

Iteration 3:
Round b0: b0→g4.
Round b1: b1→g4(b0)|b1→g5.

The sole hopeless man is **b1**. He remains coupled with **g5**.

Iteration 4:
Round b0: b0→g4.

The sole hopeless man is **b0**. He remains coupled with **g4**.

Iteration 6:
<there are no more boys remaining>
Thus the final Result = {**b0**-**g4**; **b1**-**g5**; **b2**-**g3**; **b3**-**g1**; **b4**-**g2**; **b5**-**g0**}. Notice how in this matching **b2** would get his mere 2nd preference **g3**, instead of **g1**. Notice also how girl **g2** would get her at least third preference, **b4**.

However, if only **b2** does not deviate, the outcome is as follows.

Iteration 1:
Round b0: b0→g4.
Round b1: b1→g4(b0)|b1→g5.
Round b2: b2→g1.
Round b3: b3→g1|b2→g3.
Round b4: b4→g2.
Round b5: b5→g5(b1)|b5→g0.

The hopeless men are **b2**, **b4** and **b5**. They remain coupled with **g3**, **g2** and **g0** respectively.

Iteration 2:
Round b0: b0→g4.

Round b1: b1→g4(b0) | b1→g5.

Round b3: b3→g1.

The sole hopeless man is **b3**. He remains coupled with **g1**.

Iteration 3:
Round b0: b0→g4.
Round b1: b1→g4(b0) | b1→g5.

The sole hopeless man is **b1**. He remains coupled with **g5**.

Iteration 4:
Round b0: b0→g4.

The sole hopeless man is **b0**. He remains coupled with **g4**.

Iteration 6:
<there are no more boys remaining>
Thus the final Result = {**b0**-**g4**; **b1**-**g5**; **b2**-**g3**; **b3**-**g1**; **b4**-**g2**; **b5**-**g0**}. Notice how in this matching **b2** would get his mere 2nd preference **g3**, instead of **g1**. Notice also how girl **g2** would get her at least third preference, **b4**.

Thus, **b2** and **g2** need each other to attain a better outcome for both.
|
*Discussion*: Note that in this examples, once **b2** decides to deviate, it is in the best interest of **g2** to also deviate. However, could any cases arise when if **b2** deviates and girl **g2** betrays, **g2** might get a better outcome? We leave this question open. Note also that, as in Example 3.24 (lies of b2), the actions of player **b2** are prohibited under Assumption 3.20 (no relative suicidal plays).

Advancing to the general model, we note that here it can prove even more useful for boys to have girls in their coalitions: girls might lie that they prefer one (or some) of the coalition boys over an external player (that they actually want to end up with) thus facilitating his cooption. If the external player knows that the girl is lying, the girl can still punish him in case he does not cooperate with the coalition by actually changing her preferences as she threatened. How could such a girl be coopted herself in the coalition? Well, it could be that without the help of the coalition that external boy might never propose to her: if she refuses to cooperate, some boy from the coalition might leave to the external boy a better preference for him (where he would be preferred over the external player) such that the non-cooperating girl never gets a proposal from him. New standoff situations might arise. Furthermore, since girls are allowed to be players, girls might cooperate with each other to blackmail boys into joining coalitions of their own (e.g. by threating to collectively refuse other boys - which end up occupying the non-cooperators' "higher ups" and then also collectively refuse them as well). Girls may also blackmail or coopt each other! If a grand coalition of girls forms, that coalition can force any result beyond the first round of proposal by boys (those at time index 0) under Dynamic Game Model 4.6. If they are allowed to refuse even the first proposal, then a grand coalition of girls can force its desired outcome on boys, under Dynamic Game Model 4.6. In real life situations however, in such situations boys may also be allowed to stop proposing, in which case new standoff situations might ensure between coalitions: for example, between a grand coalition of all boys versus a grand coalition of all girls; but this is not the only situation.

We leave the problem of analyzing the general model of play when girls are considered players alongside boys for future research.

Partial information about preferences

The ultimate variation which can be bought to the game involves players having to decide their strategy without having perfect knowledge about each other's preferences.

The imperfect information variation allows several situations to arise:
- Players knowing only part of the order relations in the preference lists of other boys;
- Players knowing some order relations wrongly when it comes to the preference lists of some other player.

Similar situations can arise regarding the preference lists of girls. All these situations can occur in whatever combination. Note that situations where just some order relationships in the preference list of a boy not known (with the rest being known correctly), open the possibility for that boy to lie about such. Of course, he may not know what part of his preferences is known and what part is not known. Other players might also share information (or lies) about others' preferences. Finding out the liars under certain assumptions is an interesting problem in itself and can be the object of future research.

In the case of naïve Gale-Shapely, coalitions of boys do not collectively (all players) stand to benefit from falsifying their preference lists when external players are naïve, as was shown by Dubins and Freedman in [3 - Machiavelli and the Gale-Shapley

Algorithm, L. E. Dubins and D. A. Freedman, 1981]. As discussed in Section 3 however, coalitions could improve the outcome for some of their members, not all. Same is true for the top-trading-cycles method. Furthermore, when dynamic strategy profiles are allowed, a coalition may stand to benefit by submitting falsified preference lists if it causes the external players to deviate from Gale-Shapely, as with Example 4.9 (devilish alliance). Even when only static strategy profiles are permitted, a boy might be convinced that it is useless for him to play at a certain higher-preferred girl since he would definitely (and truly) not get her anyway (as with the case of **b1** in both situations of Example 4.9).

We have already shown in Example 3.24 (lies by b2) how wrong knowledge of some player's preferences can lead to better outcomes for him in the coalition stable matching of Algorithm 3.22. We have also shown how wrong knowledge of the preference lists of both boys and girls can lead to better outcomes for them, under top-trading-cycles and coalition stable matching. The examples used there can be very easily adjusted to cases where only boys lie about their preferences: by imposing that girls actually have as preference their altered lists. This also applies to cases where their preferences are only partially known (with no incorrect knowledge), so long as the known parts are not the ones altered when lying.

In most cases players may submit static strategy profiles which are openly different from their true preference lists. Thus, they may "lie" openly. In this case other players may adjust their own strategies based on the static strategy profiles submitted (assuming they know them, of course). Players may still yet try to gain advantages by convincing a set of players that they will have a certain static strategy profile and another set that they will have a different static strategy profile, incompatible with the first, thus hoping to gain the cooperation of both groups. They may similarly lie about their true preference order.

The general play case is further complicated by wrong or imperfect knowledge of information. Some situations when girls were considered players (being able to adapt their preference lists) might as well apply in the case of imperfect knowledge of their preferences, when they are still robotic players: a boy or a coalition of boys may not know if a girl's true preferences are the ones which she would use (in the case with her being non-robotic) with him (them) cooperating or the ones she would use with him not cooperating. Thus an ambiguity might arise where priory there was a standoff. Knowing the preference lists of girls wrongly can also naturally affect outcome (for example forcing some girl-optimal matching to be played out first). Also, when the preference lists of boys are partially or incorrectly known, players might tell different lies (or the truth) to different other players so that they convince them all that they face a credible threat – or a credible promise of improvement, respectively. If the truth is not known, any of these lies could actually be the truth. Players who are desperate to avoid some outcome of a materialized threat play might accept even the possibility of the lie being truthful as a sufficient deterrent.

We leave open the subject of fully analyzing strategic play in cases of impartial information about preference lists. It can form the topic of future research.

We conclude this section here.

# 7. ECONOMIC MODELING AND APPLICATIONS

In this section we state a few real-life games and discuss how they can be modeled and resolved as instances of a stable marriage problem.

Renter – Landlord

In this problem, there are several homes which are owned by certain landlords. There are also several potential renters each of which ranks the homes on the market in some order, potentially different from the others. One renter for example might appreciate a sea-side view very much; another might appreciate a large home; and yet another might appreciate being close to his working place. Based on their appreciation of the homes and their own budgets, renters are willing to offer up until some maximum amount for renting a particular home. The landlords themselves may have a ranking over the renters, which may be based on factors such as how much they are willing to offer, weather they have pets or not, their credit rating, weather they have family and, not least, their personal feel of them.

A stable marriage problem instance readily ensues, with renters as boys (making proposals) and landlords as girls (accepting proposals).

One can see how all methods touched in this paper apply to this problem, including top-trading-cycles or the coalition stable matching of Section 3. Note the very important aspect that renters can benefits hugely from colluding, under both these methods. In fact, whenever a grand coalition ensues, the landlords get one single proposal. Thus, the renter proposing could offer a much lower budget! He no longer needs to outbid players who do not end-up with that landlord anyway. Furthermore, some players could receive side-payments from this economy (e.g. the hopeless men of prior iterations in Algorithm 3.22). The stable marriage problem modeling still holds since renters are assumed to care first and foremost about ending up in home they like. However, in such an extension (with side-payments from the economy) the last two situations described in Section 5 might arise: players could gain a preference for supporting someone who would make a higher side-payment. In cases where players care more about side-payments (above a certain threshold maybe), they may become drone players for the best paying player. Other forms of strategic plays and situations, like those discussed in Sections 4-6 might very well arise.

The problem we call "art auction" where each boy has a fixed budget he wants to spend on acquiring a certain piece of art to decorate his home and each art owner (girl) prefers the boys with the highest bid reduces to this one can also be modeled as a renter – landlord problem.

Contractor – Project

In this problem there are several contractors (either individual persons or organizations) which need to choose on which project they want to work. They may rank the available projects based on factor such as: maximum budget offered by the contracting authority; the counterparty risk of the contracting authority *with regard to them*; the required effort to gain all required skills and competences for the project; the marketing potential of the project (how likely is it to generate future great opportunities) and also personal relationships with the contracting authority. The contracting authorities may similarly rank contractors based on factors such as: the hourly / per project rate quoted by the contractor; the risks of the contractor not finishing the project in agreed terms; the location of the contractor or its employees (domestic/foreign, within the constituency of the political sponsor of the contracting authority or not, etc.); the level of security clearance of the contractor or its general trustworthiness with certain types of classified information; prospects for further benefits by engaging in a relationship with the contractor (e.g. fiscal facilities, political alliances, etc.) and, of course, personal feel. We can assume that a contractor cannot take more than one project due to business constrains: such as actual operational leverage (free operational capacity) or regulatory constraints.

Again the problem can readily can be formulated as a stable marriage problem, with either contractors or contracting authorities as boys and the others as girls. In most real-life cases it is contracting authorities who start tenders, thus assuming the role of girls.

The problem is similar to the Renter – Landlord one, with contractors as renters. One difference is that in this case renters (contractors) do not make an economy when the landlord (contracting authority) receives only their offer, but instead make a bigger profit by bidding a higher budget. Unlike in Renter – Landlord, contractors interest for a particular project might change with the budget they could attain. So might the interest of the contracting authority change with the quote of a certain contractor.
We can adapt Gale-Shapley to handle such cases: When more than one contractor proposes to the same contracting authority, the following happens:
1. The contractors which proposed to the same contracting authority (including the one which may be tentatively matched to the project) quote a project budget (this is just a simulated run – the contracting authorities are not actually informed of this; their potential responses are assumed to be known), representing the minimal amount they would be willing to accept for that project, without considering other projects first. This quote must be lower than all previous quotes he made for that project (initially it can be up to the maximal budget of the contracting authority). The contracting authority is tentatively said

   to be matched to the contractor which ranks best given these minimal bids (and considering all the other factors too, of course) – we assume that lower bids can only be better in the eyes of the contracting authority, no matter from whom.
2. The contractors which are rejected in step 1 above become uncoupled.
3. <u>Contractors may propose to a contracting authority they proposed to before, however quoting the new lowest bid (which must be necessarily lower than the prior one)</u>.

Observation 2.6 (non-decreasing temperature of girls) still holds, with properly adjusting the definition of a girl's temperature to include, besides the boy, also his bid: contracting authorities (girls) get better and better matchings in term of boy-bid tuples. Also, all notations have to be extended to include the bid amount for a particular proposal besides the girl. Also, the (tentative or final) match for a boy needs to be defined to also include, besides the partner also the bid he used to get her. Under these adjustments all other results from Section 2 and Section 3 still hold: Lemma 2.9, Lemma 3.7, Observation 3.13, Lemma 3.14, Theorem 3.21, Observation 3.28 are all still valid, by the same proofs under the adjusted notations. Algorithm 3.22 is also runnable under the modified Gale-Shapely described above. Furthermore, all examples from the usual cases still apply to cases where players also place bids: we can consider that the maximal amount of any player's bid is 0 (also the minimal).

Note that the contractors also stand to gain a lot by colluding either by top-trading-cycles or coalition-stable matching. Or by threat-making or any other form of strategic play for that matter, assuming of course it worked (produces the desired changes in the behavior of others). One clear benefit is that in a grand coalition of colluding players (as is the case with coalition-stable matching), each contracting authority gets 1 single bid for its project. Thus, in real-life play, that bid can be the maximum budget of the contracting authority.

If the contractors are allowed to make side-payments in the same currency of the bid (or transferable from such) or to subcontract a project they were awarded the problem complicates further, becoming very similar to the problem of ads placement.

The contractor – project problem is most relevant to large projects, such as those in the verticals of military, oil & gas, healthcare and to a lesser extent city management.

<u>Student - Program placement</u>

The classical problem of student-program placement can be stated as follows: there are **n** students competing for the (potentially multiple) slots of **m** University Programs (from different universities usually). Each student has a ranking preference over the programs and each University Program has a ranking preference over which students it wants to admit.

The problem can be directly formulated as a stable marriage problem – with students as boys and Universities as girls – under the variation discussed in Section 6 which allows girls to have multiple slots.

This problem is also relevant for placing primary six students in secondary schools in Singapore, which served as one motivation for [9 - Gale-Shapley Stable Marriage Problem Revisited: Strategic Issues and Applications, Chung-Piaw Teo, Jay Sethuraman,Wee-Peng Tan, 1999]) and also for placing 8$^{th}$ grade graduates in high-schools in Romania. The problem is also known as hospitals/residents problem, due to its application in the matching residents to hospitals by the non-profit organization National Resident Matching Program (NRMP) who uses the algorithm in [15 - The Redesign of the Matching Market for American Physicians: Some Engineering Aspects of Economic Design, Alvin, Elliott Peranson, 1999].

Ads placement

In the ads placement problem, there are **n** advertisers competing for **m** advertising slots. An advertising slot can be viewed as a time interval in which an advertisement is displayed in a given context (e.g. at a certain minute when a certain YouTube.com video is viewed by a user having certain characteristics – such as gender or age –, during a particular time interval – for example between 12:00 and 14:00 his local time).

Advertiser have a strict preference over the advertisement slots, based on the utility they feel they derive from the ad being delivered in the slot's context. This utility in turn may be based on factors such as: expected volume of purchases generated by the ad; branding benefits; costs of placing the ad. The advertisement slots also have a strict preference over the advertiser, based on criteria such as: amount of money the advertiser is willing to pay; the degree in which the advertiser content is expected to be relevant to viewers of the advertisement; the reputation of the advertiser; the suitability of the advertiser's content for the viewers (e.g. how much does it annoy them to have their experience interrupted by the advertisement) and so on.

This problem is in fact a variation of the Contractor – Project problem.

A case where advertisers are allowed to make side payments in the same currency as that of the bid and furthermore their utility can be valued in units of the same currency, was discussed in [13 - Designing a strategic bipartite matching market, Rahul Jain, 2007]. In this paper, Rahul Jain also noted other similar cases which were analyzed by other authors.

This problem has directly applicability in online ads placement in *Google*'s *AdSense* program as claimed in [13 - Designing a strategic bipartite matching market, Rahul Jain, 2007]

Wireless Communications

According to [16 - Matching and Cheating in Device to Device Communications Underlying Cellular Networks, Yunan Gu, Yanru Zhang, Miao Pan, Zhu Han, 2015], the stable marriage problem also has applications in optimizing the "the system throughput while simultaneously meeting the quality of service (QoS) requirements for both D2D users and cellular users (CUs)". In this paper, the authors refer to the top-trading-cycles method of [10 - Cheating by Men in the Gale-Shapley Stable Matching Algorithm, Chien-Chung Huang, 2006] to derive a "cheating matching". Unbeknownst to them is that in this paper we provided a more realistic "cheating matching", namely the coalition stable outcome of Algorithm 3.22.

We conclude this section here.

## 8. CONCLUSION AND FURTHER RESEARCH

In this paper we achieved the following:
- Provided an $O(n^2)$ algorithm which computes the coalition-stable matching under the assumption set of Section 3, about which we argued is more realistic than the one of the top-trading-cycles method used by [10 - Cheating by Men in the Gale-Shapley Stable Matching Algorithm, Chien-Chung Huang, 2006]. We have also shown a case where players can lie about their preferences to alter their outcome under this algorithm.
- Analyzed in some detail the game model for when players are allowed to make credible threats which imply potential hurtful outcomes to themselves also. We also presented super-exponential time algorithm which computes a threat-making strategy for a player **b** to get some girl **g** or better, which can, however, produces false negatives – cases where such a strategy exists but the algorithm did not find it.
- We discussed six further possibilities which might affect the outcomes for some boy, namely adding/removing boys/girls, coopting players in coalitions without incentivizing them within the game and making use of drone players. We also showed how such possibilities might become available to some player from a motivational perspective of the involved.
- We discussed in reasonable detail six major variations to the game model, namely girls with more slots, indifference in the preference lists of boys, boys not knowing about all the girls in play, boys not knowing about all other boys, girls as players and imperfect information. We provided examples to illustrate the impact such variations have on prior results.
- We presented four real life games namely Renter-Landlord, Contractor – Project, Student – Program placement and Ads placement and analyzed how they can be modeled as stable marriage problem instances. In the course of doing this we also expanded the model to allow a seventh major variation, as discussed in Section 7 regarding the contractor – project problem. We also referenced a fifth game, namely Wireless Communications where our results also apply.

Throughout this paper we have left several open questions and topics for future research. They are summarized below:

- Can the matching produced by Algorithm 3.22 be computed faster than O($n^3$)?
- Is there an algorithm more efficient in terms of running time than walking the entire game DAG for determining **iff** there exists a resolution to standoffs which make a certain boy **b** certain to get some girl **g** or better under the assumption set of Section 4? What further assumptions would such an algorithm need to make?
- Can the output of Algorithm 4.26 (expanding wrath approach) – namely the permutation under which a coalition expands – be computed by a more efficient algorithm in terms of running time?
- For an augmented problem instance, is there an efficient algorithm for solving the game fully? What assumption does it need to make about standoffs resolution?
- In case players are allowed to lie or not respond to agreement offers, how can this impact the algorithms presented?
- Questions pertaining to situations when girls are players:
  - Is there an efficient algorithm for determining potential coalitions of girls, along with associated outcomes for when boys are assumed to be robotic naïve Gale-Shapely players?
  - What is an appropriate model for describing situations when girls are allowed to be players and form or join coalitions of other girls and boys? Is there an efficient algorithm for computing weather there exists a resolution to stand-offs such that a boy **b** gets girl **g** or better or, conversely, that girl **g1** gets boy **b1** or better under this model?
- Question pertaining to partial information about preferences of other players:
  - When players can divulge (or lie about) some order relationships in their preference list to other players, how can liars be found out? Is there an efficient algorithm which determines that a player is lying? We strongly suspect a positive answer.
  - How can players speculate the lack of knowledge (or wrongful knowledge) of other players? Is there an efficient algorithm for computing an optimal (or at least a good) strategy for each player to make use of such shortcomings of others? What assumptions must it make?

We conclude the paper here.